\documentclass[twocolumn,aps,floatfix,superscriptaddress,longbibliography]{revtex4-1}

\usepackage[hidelinks,colorlinks=true,citecolor=blue]{hyperref}
\usepackage{float}
\usepackage{upgreek}
\usepackage{graphicx} 
\usepackage{gensymb}
\usepackage{epsfig}
\usepackage{amsmath,bm,amssymb}
\usepackage{amsfonts}
\usepackage[usenames]{color}
\usepackage{times}

\begin{document}

\title{Frozen spin ratio and the detection of Hund correlations}

 \author{Siheon Ryee}
 \email{sryee@physnet.uni-hamburg.de}
 \affiliation{Department of Physics, KAIST, Daejeon 34141, Republic of Korea}
 \affiliation{I. Institute of Theoretical Physics, University of Hamburg, Notkestrasse 9, 22607 Hamburg, Germany}

 \author{Sangkook Choi}
 \email{sangkookchoi@kias.re.kr}
 \affiliation{Condensed Matter Physics and Materials Science Department, Brookhaven National Laboratory, Upton, NY 11973, USA}
 \affiliation{School of Computational Sciences, Korea Institute for Advanced Study, Seoul 02455, Republic of Korea}

 \author{Myung Joon Han}
 \email{mj.han@kaist.ac.kr}
 \affiliation{Department of Physics, KAIST, Daejeon 34141, Republic of Korea}

\date{\today}

\begin{abstract}	
We propose a way to identify strongly Hund-correlated materials by unveiling a key signature of Hund correlations at the two-particle level. The defining feature is the {\it sign} of the response of the {\it frozen spin ratio} (the long-time local spin-spin correlation function divided by the instantaneous value) under variation of electron density. 
The underlying physical reason is that the sign is closely related to the strength of charge fluctuations between the dominant atomic multiplets and higher-spin ones in a neighboring charge subspace. It is the predominance of these fluctuations that promotes Hund metallicity. The temperature dependence of the frozen spin ratio can further reveal a non-Fermi-liquid behavior and thus the Hund metal states.  
We analyze both degenerate and non-degenerate multiorbital Hubbard models and corroborate our argument by taking doped La$_2$CuO$_4$ and LaFeAsO as representative material examples, respectively, of Mott and Hund metals.  Our proposal should be applicable to systems with non-half-filled integer electron fillings and their doped cases provided the doping drove the electron density toward the half filling.
\end{abstract}
 
\maketitle

\section{Introduction}

Understanding physical properties of a given system is largely dictated by a ``reference frame" that inherits the characteristics of a relevant physical picture or a simple model.
In strongly correlated materials, bad metal behavior has commonly been associated with Mott physics. In such a frame, large effective Coulomb repulsion is responsible for slowing down electron motion by penalizing double occupancy of electrons on the same site \cite{Imada}. 
A widely-accepted material example pertaining to this category is cuprates which display several intriguing phases including superconductivity as a function of doping \cite{PALee,Norman,Keimer}.

In some multiorbital materials such as ruthenates and Fe-based superconductors (FeSCs), on the other hand, the nature of their correlated metallic phases is far from the conventional Mott paradigm  \cite{Haule,Mravlje,Werner3}.
In this respect, it has been emphasized over many years that Hund coupling $J$ is a new route to strong correlations with the dawn of the concept, Hund metal \cite{Werner1,Haule,Nevidomskyy,Mravlje,Werner3,Medici_2011,Janus,Yin1,Yin_power,Toschi,Georges,Medici_2014,Khajetoorians,Fanfarillo,Hoshino,SOS1,Stadler1,SOS2,Stadler2,Deng,Isidori,Ryee,Huang_kagome,Chen_ferromagnet,HJLee2,YWang,Watzenbock,Karp,Fanfarillo_2,BKang,Bramberger,Ryee3,Drouin,Stadler3,HJLee,Stepanov,Nomura,Drouin_2022,TJKim}. A key notion here is that a sizable $J$ impedes the formation of long-lived quasiparticles by suppressing the screening of local spin moments  \cite{Nevidomskyy,Georges,Yin1,Yin_power,SOS1,SOS2,Stadler1,Stadler2,Deng,Ryee3,Drouin,Stadler3,Drouin_2022}.

A central question at this stage is the following: What are the hallmarks of Hund physics distinctive from Mott physics? The question also concerns whether they can be experimentally measurable. The difficulty lies in the fact that strong correlations in multiorbital materials cannot solely be attributed to $J$; in some sense, the influences of $J$ and $U$ are intertwined with each other \cite{Janus,Georges}. Furthermore, the additional energy scales (e.g., crystal-field splitting) add more complexity to this problem.

\begin{figure*} [!htbp] 
	\includegraphics[width=0.99\textwidth]{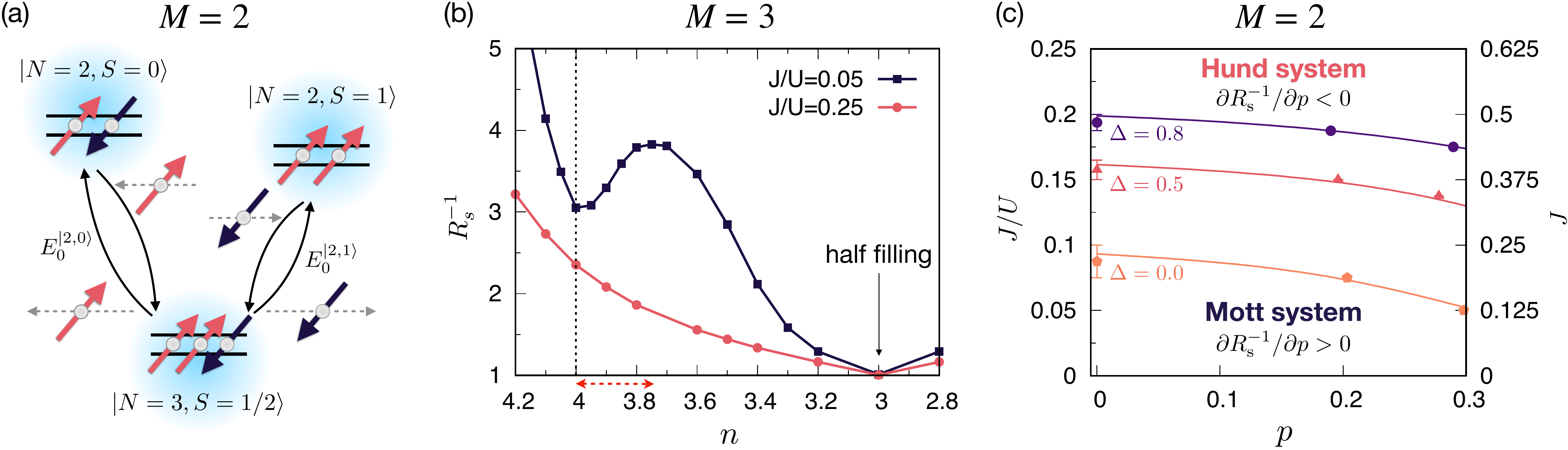}
	\caption{(a) Schematic illustration of two different charge fluctuations on a site having $|N=3,S=1/2 \rangle$ as the atomic ground state multiplet when $M=2$. Non-Hund fluctuations (left paths with an energy cost $E_0^{|2,0\rangle}$ forming singlet $|N=2, S=0\rangle$) and Hund fluctuations (right paths with an energy cost $E_0^{|2,1\rangle}$ forming triplet $|N=2, S=1\rangle$) are depicted. Colored arrows represent spin moments. (b) $R_\mathrm{s}^{-1}$ plotted against decreasing electron density $n$ for a three-degenerate-orbital ($M=3$) model at $U=3.5$.  
	The red horizontal double arrow below the $x$-axis indicates a range of filling in which $\partial R_\mathrm{s}^{-1}/\partial p>0$ for $J/U=0.05$ and $\partial R_\mathrm{s}^{-1}/\partial p<0$ for $J/U=0.25$. The system is particle-hole symmetric about $n=3$.
	(c) The diagram differentiating Mott and Hund systems for $M=2$ and $U=2.5$. Here, $n=M+1-p=3-p$. The solid lines with symbols denote the region above (below) which $\partial R_\mathrm{s}^{-1}/\partial p<0$ ($\partial R_\mathrm{s}^{-1}/\partial p>0$) for three different values of $\Delta$. 
    }
	\label{fig1}
\end{figure*}

In this paper, we identify direct manifestations of Hund correlations at the two-particle level. To this end, we propose a two-particle quantity, which we call the frozen spin ratio, $R_\mathrm{s}$. It is defined as the long-time local spin-spin correlation function divided by the instantaneous value [see Eq.~(\ref{eq2}) below]. Specifically, we argue that the {\it sign} of the response of $R_\mathrm{s}^{-1}$ under variation of electron density $n$ is the key defining feature to classify two regimes of Mott and Hund correlations in the most relevant parameter range. The underlying physical reason is that the sign is closely related to the strength of ``Hund fluctuations": ferromagnetic charge fluctuations between the dominant atomic multiplets and higher-spin ones in a neighboring charge subspace [see Fig.~\ref{fig1}(a)]. It is the predominance of these fluctuations that impedes screening of local spin moments and promotes Hund metallicity \cite{Yin_power,SOS2,Georges}.

\section{Models and Method}

We consider $M$-orbital ($M \ge 2$) Hubbard models with onsite Coulomb interactions. The local part of our Hamiltonian is given by
\begin{align}
H_\mathrm{loc} &=  U\sum_{\eta}{n_{\eta \uparrow} n_{\eta \downarrow}} 
+ \sum_{\eta < \eta',\sigma\sigma'}(U'-J\delta_{\sigma\sigma'}){ n_{ \eta \sigma} n_{ \eta' \sigma'}} \nonumber \\
&+ J\sum_{\eta \neq \eta'}(d^{\dagger}_{\eta \uparrow}  d^{\dagger}_{\eta' \downarrow} d_{\eta \downarrow} d_{\eta' \uparrow} 
+d^{\dagger}_{ \eta \uparrow} d^{\dagger}_{ \eta \downarrow} d_{ \eta' \downarrow} d_{ \eta' \uparrow}) \nonumber \\
& +\sum_{\eta,\sigma}(\epsilon_\eta - \mu)n_{\eta\sigma},
\label{eq1}
\end{align}
where $d^{\dagger}_{\eta \sigma}$ ($d_{\eta \sigma}$) is the electron creation (annihilation) operator for orbital $\eta,\eta' \in  \{ 1, 2,...,M \}$ and spin $\sigma,\sigma' \in \{ \uparrow, \downarrow \}$. $n_{\eta\sigma} = d^{\dagger}_{\eta \sigma}d_{\eta \sigma}$ is the electron number operator. $\epsilon_\eta$ is the onsite energy level of orbital $\eta$ and $\mu$ is the chemical potential. $U$ ($U'$) denotes the intraorbital (interorbital) Coulomb repulsion. We set $U'=U-2J$ following the convention. 
For the kinetic part of our Hamiltonian, we mainly consider an infinite-dimensional Bethe lattice with an equal half-bandwidth $D$ for each nonhybridized orbital. $D$ is used as the energy unit. As a proof of concept, we present results for La$_2$CuO$_4$ and LaFeAsO using {\it ab initio} model parameters. The models are solved using the dynamical mean-field theory (DMFT) \cite{DMFT,Kotliar} by employing the hybridization-expansion continuous-time quantum Monte Carlo algorithm as an impurity solver \cite{CTQMC,Choi}.

\section{Mott versus Hund systems in terms of the frozen spin ratio}
\subsection{Frozen spin ratio $R_\mathrm{s}$}
The central quantity we investigate is the frozen spin ratio $R_\mathrm{s}$ which is defined as
\begin{align}
R_\mathrm{s} \equiv { C(\tau=\frac{1}{2T}) }/{ C(\tau=0) },
\label{eq2} 
\end{align}
where $C(\tau) \equiv  \langle S_z(\tau) S_z(0) \rangle$ is the local spin-spin correlation function with
$S_z(\tau) = \sum_{\eta}n_{\eta\uparrow}(\tau)-n_{\eta\downarrow}(\tau)$ ($\tau$: imaginary time; $\tau \geq 0$) being the local spin operator. $T$ is temperature. 
On general grounds, $C(\tau)$ decays over $\tau$ ($\tau \leq 1/(2T)$) due to the dynamical nature of spin moments. For an isolated atom having local moments, however, $R_\mathrm{s}=1$ because $C(\tau) = C(0)$ for the entire range of $\tau$. In a Fermi-liquid (FL) limit where the dynamical screening becomes very effective, on the other hand, $C(\tau) \rightarrow 0$ at long times $\tau$ (or $T \rightarrow 0$), namely at $\tau \gg 1/T_\mathrm{K}$ ($T_\mathrm{K}$: the Kondo temperature below which the local moments are screened and long-lived quasiparticles are formed) \cite{Stadler2,Kowalski}, resulting in $R_\mathrm{s} \rightarrow 0$. Thus $R_\mathrm{s}$ should take an appreciable value well above $T_\mathrm{K}$. We will examine $R_\mathrm{s}$ at $T=0.02$, unless otherwise specified.

Note that $R_\mathrm{s}$ is a measure of the degree of spin screening, not the magnitude of local spin moments, and is normalized; $R_\mathrm{s} \in [0,1]$ lying in between two extreme limits of a low-temperature fully-screened regime ($R_\mathrm{s} \rightarrow 0$) and the unscreened local moment ($R_\mathrm{s} \rightarrow 1$). 
We will also use $R_\mathrm{s}^{-1}$ as well as $R_\mathrm{s}$.
Hence, $R_\mathrm{s}^{-1}$ is reduced down toward unity (not zero) as the system moves toward the local moment regime.

\subsection{Identification of Mott and Hund systems}

We begin with the archetypal case: a three-degenerate-orbital model ($M=3$ and $\epsilon_1=\epsilon_2=\epsilon_3$). Figure~\ref{fig1}(b) presents $R_\mathrm{s}^{-1}$ plotted against decreasing electron density $n$ for two different values of $J/U$. Most notably, one can identify that, near the typical ``Hund-metal electron filling" $n=M + 1$ \cite{Georges}, $R_\mathrm{s}^{-1}$ exhibits a ``V-shape" behavior in the small $J/U$ case, whereas it is monotonically decreasing for the large $J/U$. Thus, while $\partial R_\mathrm{s}^{-1} / \partial p>0$ for small $J/U$, $\partial R_\mathrm{s}^{-1} / \partial p<0$ for large $J/U$ by hole doping $p$ to $n=M+1$.  The same behavior occurs for $n=M-1$ provided the electron is doped (not shown).  In general, the dichotomy of $R_\mathrm{s}^{-1}$ can also emerge for any non-half-filled integer electron fillings. Throughout the paper we will focus on ``near $M+1$" densities, namely $n=M+1-p$ ($0 \leq p <0.5$). Note that the dichotomy between small and large $J/U$ regimes does not appear for $p<0$ or large $p$ ($p \rightarrow 1$) by which the system gets too close to the half filling as shown in Fig.~\ref{fig1}(b). Thus, our proposal cannot be applicable to these cases; more discussion can be found in Appendix \ref{appendix_F} on the applicability of $\partial R_\mathrm{s}^{-1} / \partial p$ in identifying Hund systems.

The behavior of $R_\mathrm{s}^{-1}$ in the small $J/U$ case can be naturally understood from Mott physics by which the correlation strength gets mitigated as the system moves away from an integer filling. As a consequence, Kondo screening of local spin becomes more effective (i.e., enhancement of $R_\mathrm{s}^{-1}$) by doping (either electron or hole) an integer-filled system as shown in Fig.~\ref{fig1}(b). 
On the other hand, this picture fails to explain the observed behavior of $R_\mathrm{s}^{-1}$ in the large $J/U$ case. 
Furthermore, since $J$ drives the system away from a Mott insulator at $n=M+1$ \cite{Georges}, we infer that strong Hund correlations distinctive from Mott physics are manifested by $\partial R_\mathrm{s}^{-1} / \partial p<0$.
Importantly, this characteristic feature is identified in various other cases; see Appendices \ref{appendix_A}--\ref{appendix_C}.  ``Mott systems" hereafter refer to not only Mott insulators but also correlated metals governed by Mott physics. Note that we do not focus on the question of how correlated the system is in this paper. In order to do so, one can investigate the well-established physical quantities such as quasiparticle weight, spectral functions, and dc/optical conductivities.

To highlight the generality of the above observation, we present a diagram in Fig.~\ref{fig1}(c) indicating the regions of Mott and Hund physics dominant correlations as determined by the sign of $\partial R_\mathrm{s}^{-1} / \partial p$ for a two-orbital ($M=2$) model  using three different values of on-site energy-level splitting $\Delta$ ($\Delta \equiv \epsilon_1-\epsilon_2$). 
Figure~\ref{fig1}(c) demonstrates that the characteristic of Mott systems ($\partial R_\mathrm{s}^{-1} / \partial p > 0$) becomes pronounced for small $J/U$ and large $\Delta$, whereas that putatively of the Hund system ($\partial R_\mathrm{s}^{-1} / \partial p < 0$) becomes pronounced for large $J/U$ and small $\Delta$. This behavior is consistent with the common notion that $\Delta$ suppresses Hund physics.

A useful insight can be obtained by examining the Kondo coupling for spin, namely $\mathcal{J}_\mathrm{s}$, as derived from the Schrieffer-Wolff (SW) transformation \cite{SWT,Yin_power,SOS1,SOS2}. 
While various $\mathcal{J}_i$s are coupled through scaling equations, by neglecting cross terms between Kondo couplings it is only $\mathcal{J}_\mathrm{s}$ (Kondo coupling for spin) that determines $T_\mathrm{K}$ \cite{SOS1,SOS2}. With this idea in mind, we argue that the observed behavior, namely the negative response of $R_\mathrm{s}^{-1}$ (i.e., $\partial R_\mathrm{s}^{-1} / \partial p < 0$) features strong Hund fluctuations which reduce $\mathcal{J}_\mathrm{s}$ and $T_\mathrm{K}$.

\begin{figure} [!htbp] 
	\includegraphics[width=0.95\columnwidth]{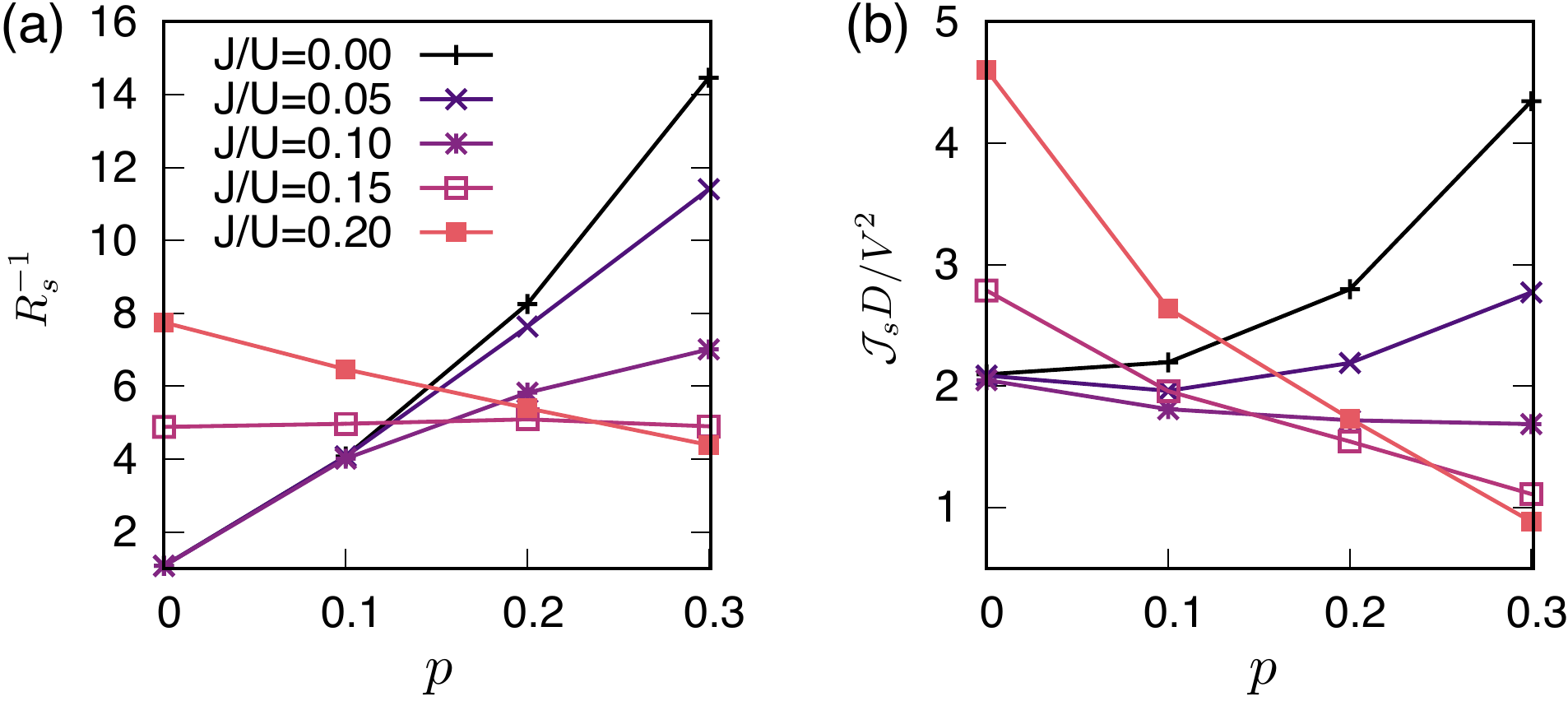}
	\caption{Hole doping dependence of (a) $R_\mathrm{s}^{-1}$ and (b) $\mathcal{J}_\mathrm{s}$ (multiplied by $D/V^2$) for $M=2$, $\Delta=0.5$, and $U=2.5$. The electron density $n$ is given by $n=3-p$. We used $E_k^{|N,S\rangle}$s obtained from the DMFT solutions for the evaluation of $\mathcal{J}_\mathrm{s}$. 
	}
	\label{fig2}
\end{figure}

Let us examine $\mathcal{J}_\mathrm{s}$ for $M=2$ which roughly reads $\mathcal{J}_\mathrm{s} \sim \sum_{k,N\in\{2,4\},S}\mathcal{O}(V^2/E_k^{|N,S\rangle})$; see Appendix \ref{appendix_D} for the derivation. Here, $V$ is the bath-impurity hybridization strength. $E_k^{|N,S\rangle}$ ($E_k^{|N,S\rangle}>0$) denotes an excitation energy from $|3,1/2\rangle $ to the excited atomic multiplet $|N,S \rangle$. The subscript $k$ ($k \in \{0,1,...\}$) refers to the $k$-th lowest eigenvalue in the corresponding $|N,S\rangle$ subspace. Figures~\ref{fig2}(a--b) present $R_\mathrm{s}^{-1}$ and $\mathcal{J}_\mathrm{s}$ as a function of $p$. Interestingly, we find qualitatively the same behavior of $R_\mathrm{s}^{-1}$ and $\mathcal{J}_\mathrm{s}$; see Appendix~\ref{appendix_E} for more data.

To understand the above observation, we consider the response of $\mathcal{J}_\mathrm{s}$ upon density change. To mimic the effect of a small increase in $p$, let us consider a situation where $\mu$ is slightly decreased by $d\mu$ ($d\mu>0$), i.e., $\mu \rightarrow \mu- d\mu$, which leads to $\mathcal{J}_\mathrm{s} \rightarrow \mathcal{J}_\mathrm{s} - ( \partial \mathcal{J}_\mathrm{s} / \partial \mu ) d\mu$ to the first order.
The leading contribution to the change in $\mathcal{J}_\mathrm{s}$, namely $- ( \partial \mathcal{J}_\mathrm{s} / \partial \mu ) d\mu$, is given by (Appendix \ref{appendix_D}):
\begin{align}
-\Big( \frac{\partial \mathcal{J}_\mathrm{s}}{\partial \mu} \Big) d\mu \approx -\frac{1}{2} \frac{V^2 d\mu}{ \big(E_0^{|2,1\rangle} \big)^2} + f(\Delta/J) \frac{V^2 d\mu}{ \big(E_0^{|2,0\rangle} \big)^2},
\label{eq3}
\end{align}
where $f(\Delta/J) = 1/[\{ \sqrt{1+(\Delta/J)^2} -\Delta/J\}^2+1]$ and $0<f(\Delta/J)<1$. $E_0^{|2,1\rangle} - E_0^{|2,0\rangle} = \sqrt{J^2+\Delta^2}-3J$; refer to Table~\ref{table_s2}. Here we emphasize that the first term associated with Hund fluctuations [Fig.~\ref{fig1}(a)] is negative definite, thereby playing a major role in the decrease of $\mathcal{J}_\mathrm{s}$ by hole doping. Thus predominance of the first term suppresses spin-Kondo screening by reducing $\mathcal{J}_\mathrm{s}$, which in turn, enhances the Hund metallicity. On the other hand, the second term, which is positive, features the effect of $\Delta$ and promotes the screening. Since the second term is enhanced by $\Delta$, it can be seen that large $\Delta$ masks the effect of Hund fluctuations, which is qualitatively consistent with what we have seen in Fig.~\ref{fig1}(c) and Fig.~\ref{fig2}. The role of $J$ here is two-fold: $J$ i) drives the system away from a Mott insulator \cite{Georges} and ii) enhances the Hund fluctuations by reducing $E_0^{|2,1\rangle}$ compared to $E_0^{|2,0\rangle}$ \cite{Ryee3}. Therefore, the decrease of $\mathcal{J}_\mathrm{s}$ upon density change and concomitantly suppressed spin-Kondo screening is a genuine effect of $J$, not by the proximity to a Mott insulator.

In contrast to $\mathcal{J}_\mathrm{s}$, however, other relevant couplings in generic $\mathrm{SU}(2) \otimes \mathrm{SU}(M)$ models increase with $p$ \cite{SOS2}, evolving in such a way to promote Kondo screening as detailed in Appendix \ref{appendix_D}. We thus ascribe the negative response of $R_\mathrm{s}^{-1}$ in Hund systems to the effect of strong Hund fluctuations.

It should be noted, at this point, that the doping dependence of $R_\mathrm{s}^{-1}$ does not tell whether a system at a given temperature is a non-Fermi liquid. It only indicates whether the electron correlation of the system is governed by Hund physics or not. This limitation leads us to further look at the temperature dependence of $R_\mathrm{s}$ to uncover the nature of a metallic state as detailed below.

\subsection{Non-Fermi-liquid behavior}

\begin{figure} [!htbp] 
	\includegraphics[width=0.9\columnwidth, angle=0]{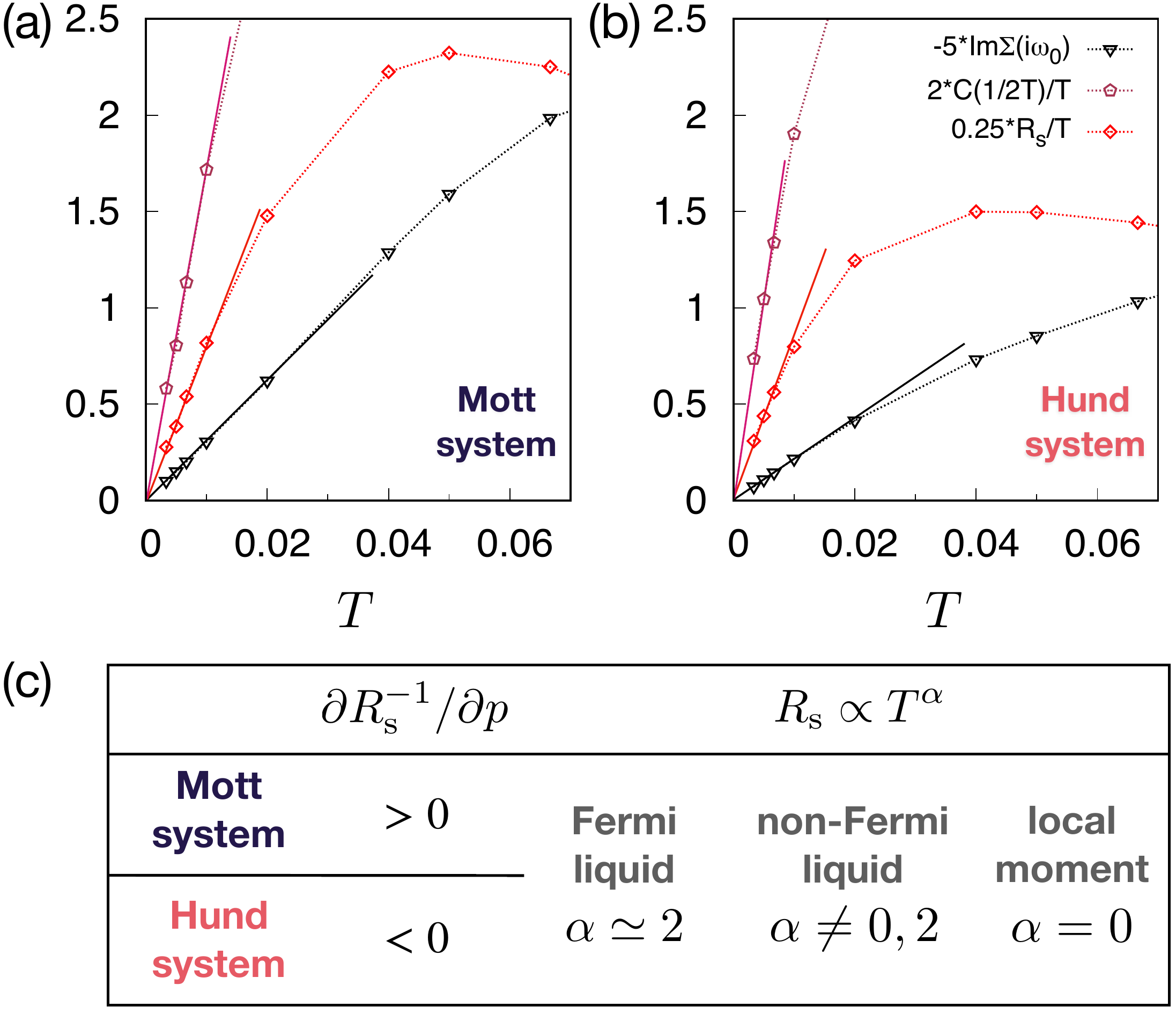}
	\caption{(a--b) Temperature dependence of $\mathrm{Im}\Sigma(i\omega_0)$, $C(1/2T)/T$, and $R_\mathrm{s}/T$ obtained from a two-degenerate-orbital model of $n=2.9$ for (a) a Mott system ($U=3$ and $J/U=0.05$; $\partial R_\mathrm{s}^{-1}/ \partial p>0$) and (b) a Hund system ($U=3$ and $J/U=0.25$; $\partial R_\mathrm{s}^{-1} / \partial p<0$). The solid lines are guides to the eye to indicate the FL behavior. (c) Characteristic features of Mott and Hund systems at low temperatures for the systems with $n=M+N-p$ ($0\leq p < 0.5$) electron densities, where $N$ is a non-zero positive integer. These features also hold for the cases of $n=M-N+p$. 
	}
	\label{fig3}
\end{figure}

Having established that the sign of $\partial R_\mathrm{s}^{-1} / \partial p$ is a useful indicator to identify Hund systems, we monitor the temperature dependence of $R_\mathrm{s}$ to tell whether a given system at a low temperature follows the FL behavior. In the FL regime at low temperatures, $C(1/2T)$ scales as $T^2$~\cite{DMFT,Werner1,Cha,Dumitrescu} and the instantaneous value, $C(0)$, becomes basically $T$-independent. Thus, $R_\mathrm{s}$ should concomitantly exhibit the $T^2$ dependence. In contrast, in a local moment regime or a Mott insulator, $R_\mathrm{s}$ is $T$-independent. 
To illustrate this argument, we look at the $T$ dependence of $R_\mathrm{s}$ divided by $T$ ($R_\mathrm{s}/T$) presented in Figs.~\ref{fig3}(a--b) for both Mott and Hund systems. We also look at the imaginary part of the local self-energy at the lowest Matsubara frequency, $\mathrm{Im}\Sigma(i\omega_0)$, which exhibits $T$-linear scaling in the FL regime as demonstrated in Ref.~\cite{Chubukov}.

We find from Figs.~\ref{fig3}(a--b) that both Mott and Hund systems exhibit the FL behavior, namely $\mathrm{Im}\Sigma(i\omega_0) \propto T$ \cite{Chubukov}, at low temperatures. Interestingly, indeed, $C(1/2T)/T \propto T$ and $R_\mathrm{s}/T \propto T$ (or, equivalently $R_\mathrm{s} \propto T^2$) below which the FL behavior sets in. 
We note that, a Hund metal, if one defines this state as a non-FL, can be identified as a Hund system ($\partial R_\mathrm{s}^{-1} / \partial p <0$) lying in a regime in which the exponent $\alpha$ of $R_\mathrm{s} \propto T^\alpha$ is close to neither 0 nor 2. A summary of our analysis so far is presented in Fig.~\ref{fig3}(c).

\subsection{Relations of $R_\mathrm{s}$ to other quantities: the quasiparticle weight and the spin-orbital separation}

\begin{figure} [!htbp]
	\includegraphics[width=1.0\columnwidth, angle=0]{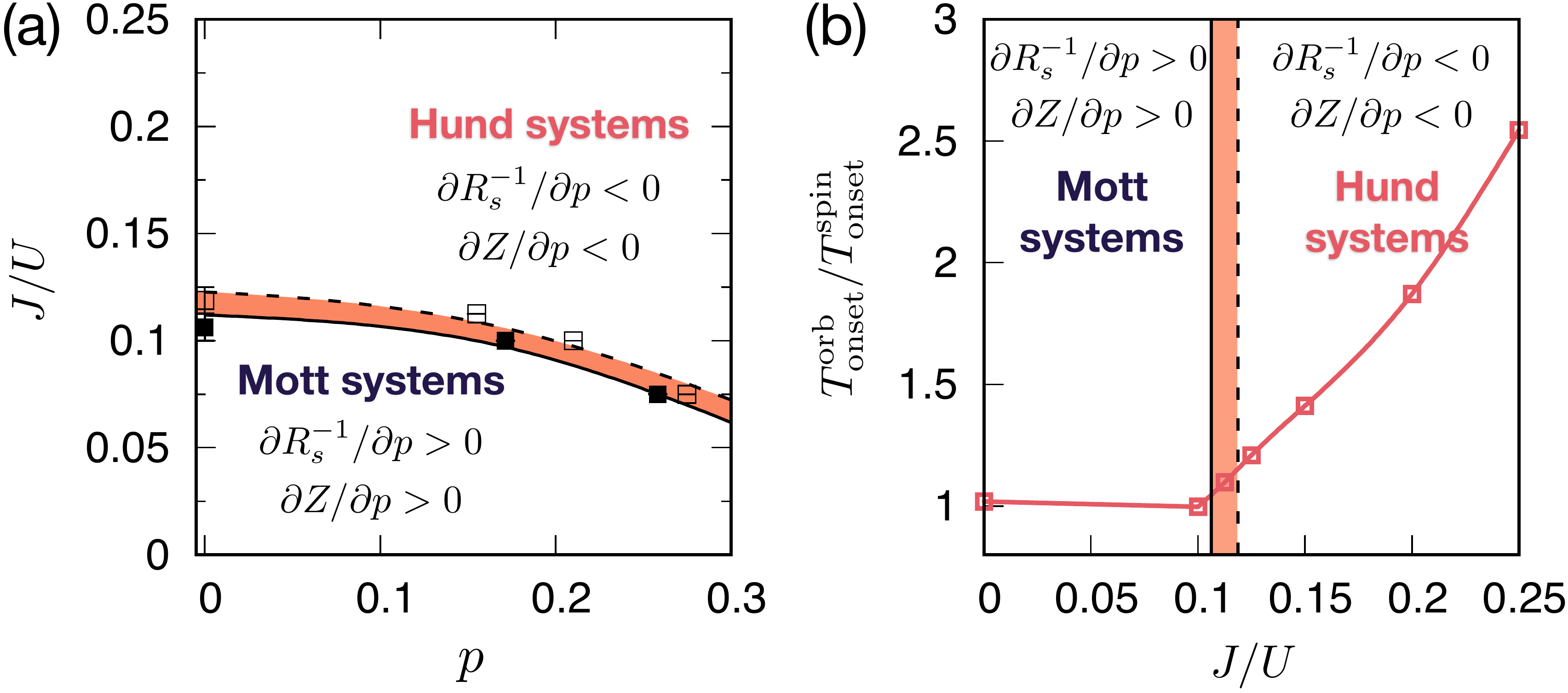}
	\caption{(a) The diagram differentiating Mott and Hund systems for a two-degenerate-orbital ($M=2$) model. $U=3$ and $n=M+1-p=3-p$. The solid (dashed) line with filled (empty) squares denotes the region above which $\partial R_\mathrm{s}^{-1}/\partial p<0$ ($\partial Z/\partial p<0$). The narrow orange area where $\partial R_\mathrm{s}^{-1}/\partial p<0$ and $\partial Z/\partial p>0$ is an intermediate region between Mott and Hund systems. We used a lower simulation temperature for the calculations of $Z$ ($T=0.005$) than for $R_\mathrm{s}^{-1}$ ($T=0.02$). (b) The ratio of $T^\mathrm{orb}_\mathrm{onset}$ to $T^\mathrm{spin}_\mathrm{onset}$ as a function of $J/U$ for the same model at $p=0$. The vertical solid (dashed) line denotes the ``Mott--Hund boundary" for $p=0$ determined by the sign of $\partial R_\mathrm{s}^{-1}/\partial p$ ($\partial Z/\partial p$).
	}
	\label{fig4}
\end{figure}

Since $T_\mathrm{K} \sim Z$ ($Z$: the quasiparticle weight or the inverse of the quasiparticle-mass enhancement within DMFT) \cite{Stadler2}, the behavior of $\partial Z / \partial p$ should follow $\partial R_\mathrm{s}^{-1} / \partial p$ in degenerate-orbital systems. Indeed, the boundary determined by $\partial Z / \partial p$ below the FL coherence temperature is almost the same as that determined by $\partial R_\mathrm{s}^{-1} / \partial p$ at intermediate temperatures as presented in Fig.~\ref{fig4}(a). Note also that $Z$ corresponds to the {\it inverse} of the mass enhancement $m_\mathrm{b}/m^*$ ($m_\mathrm{b}$ is the noninteracting band mass) within the DMFT.

We now compare the Mott--Hund boundary based on the signs of $\partial R_\mathrm{s}^{-1} / \partial p$ and $\partial Z / \partial p$ with that obtained by the onset temperatures of screening of orbital and spin degrees of freedom. These two temperatures, namely $T^\mathrm{onset}_\mathrm{orb}$ for orbital and $T^\mathrm{onset}_\mathrm{spin}$ for spin, are defined as the temperatures below which the Curie law of local moments starts to get violated and Kondo screening sets in \cite{Deng}. Strong Hund physics in (nearly) degenerate-orbital models at $p=0$ is captured by the separation of these two temperatures, namely $T^\mathrm{onset}_\mathrm{orb} > T^\mathrm{onset}_\mathrm{spin}$ \cite{Deng,Stadler2,Stadler3,Ryee3}. With this idea in mind, we here monitor the ratio $T_\mathrm{onset}^\mathrm{orb} / T_\mathrm{onset}^\mathrm{spin}$ as a function of $J/U$ for a two-degenerate-orbital ($M=2$) model at $n=M+1=3$ and $U=3$. Specifically, we first evaluate the local orbital/spin susceptibilities: $\chi_\mathrm{o/s}=\int_{0}^{1/T}{d\tau \{ \langle {O}_\mathrm{o/s}(\tau){O}_\mathrm{o/s}(0) \rangle } - \langle {O}_\mathrm{o/s} \rangle^2 \}$, where  ${O}_\mathrm{o}(\tau)=\sum_{\sigma}n_{1\sigma}(\tau)-n_{2\sigma}(\tau)$  for orbital and ${O}_\mathrm{s}(\tau)=\sum_{\eta }n_{ \eta \uparrow}(\tau)-n_{\eta \downarrow}(\tau)$ for spin. Then, we estimate $T_\mathrm{onset}^\mathrm{orb/spin}$ for each $J/U$ value by fitting the high-$T$ data to the following formula: $\chi_\mathrm{o/s} \propto 1/(T+T_\mathrm{onset}^\mathrm{orb/spin})$. Our result is presented in Fig.~\ref{fig4}(b).

Interestingly, we find that the boundary based on the signs of $\partial R_\mathrm{s}^{-1} / \partial p$ and $\partial Z / \partial p$ is in good agreement with that by $T_\mathrm{onset}^\mathrm{orb} / T_\mathrm{onset}^\mathrm{spin}$ [Fig.~\ref{fig4}(b)]. We can understand it by investigating the Kondo couplings.
In the regime of $J \ll U$, $\mathcal{J}_i \simeq \mathcal{J}^{J=0}_i$ because $\mathcal{J}_{i}/V^2 = \mathcal{J}^{J=0}_i/V^2 + \mathcal{O}({J}/{U^2}) + \mathcal{O}({J^2}/{U^3}) + \cdots$. Here, $\mathcal{J}^{J=0}_i$ refers to the Kondo couplings when $J/U=0$. In this case, the relation $\mathcal{J}_\mathrm{o} \simeq \mathcal{J}_\mathrm{s}$ ($\mathcal{J}_\mathrm{o}$: Kondo coupling for orbital) holds under renormalization group flow, whereby $T_\mathrm{onset}^\mathrm{orb} / T_\mathrm{onset}^\mathrm{spin} \simeq 1$ as discussed in Ref.~\cite{Ryee3}. Notably, since $J \ll U$ here, Hund fluctuations which give rise to $\partial R_\mathrm{s}^{-1} / \partial p<0$ and $\partial Z / \partial p<0$ are largely masked by non-Hund fluctuations which favor $\partial R_\mathrm{s}^{-1} / \partial p>0$ and $\partial Z / \partial p>0$. Thus, for degenerate-orbital models at $p \rightarrow 0$, we obtain $\partial R_\mathrm{s}^{-1} / \partial p>0$ and $\partial Z / \partial p>0$ for the same $J/U$ range in which $T_\mathrm{onset}^\mathrm{orb} / T_\mathrm{onset}^\mathrm{spin} \simeq 1$. Note, however, that the applicability of our criterion based on the sign of $\partial R_\mathrm{s}^{-1} / \partial p$ is not limited to degenerate-orbital models at integer fillings.  Thus, the criterion is particularly useful in situations where multiple partially-filled orbitals are non-degenerate.

\section{Proof of a concept: cuprates and Fe-based superconductors}

To corroborate the validity of our proposal, we address two representative Mott and Hund materials. To this end, we solve a two-orbital model ($e_g$ orbitals; $M=2$) for La$_2$CuO$_4$ and a five-orbital model ($M=5$) for LaFeAsO using {\it ab initio} parameters.  
For La$_2$CuO$_4$, we use parameters for a two-orbital model as derived in Ref.~\cite{Hirayama}; see Table~1 and Table~S15 in Ref.~\cite{Hirayama}. We take only onsite Coulomb interactions for the model. For LaFeAsO, we use in-plane hopping amplitudes listed in Table~IV in Ref.~\cite{Miyake}. For two-body terms, $U$, $U'$, and $J$ values of $U=2.53$, $U'=1.756$, and $J=0.387$~eV are adopted for Eq.~(\ref{eq1}) by parametrizing onsite Coulomb interaction elements listed in Table~VIII of Ref.~\cite{Miyake}. We take two-dimensional lattices using in-plane hoppings for both systems and use $80 \times 80$ $\mathbf{k}$ points in the first Brillouin zone.
In their undoped ($p=0$) forms, they both possess $M+1$ electrons. Our main interest is the filling dependence of the inverse of the quasiparticle-mass enhancement $m_\mathrm{b}/m^*$ and $R_\mathrm{s}^{-1}$.

\begin{figure} [!htbp] 
	\includegraphics[width=0.93\columnwidth]{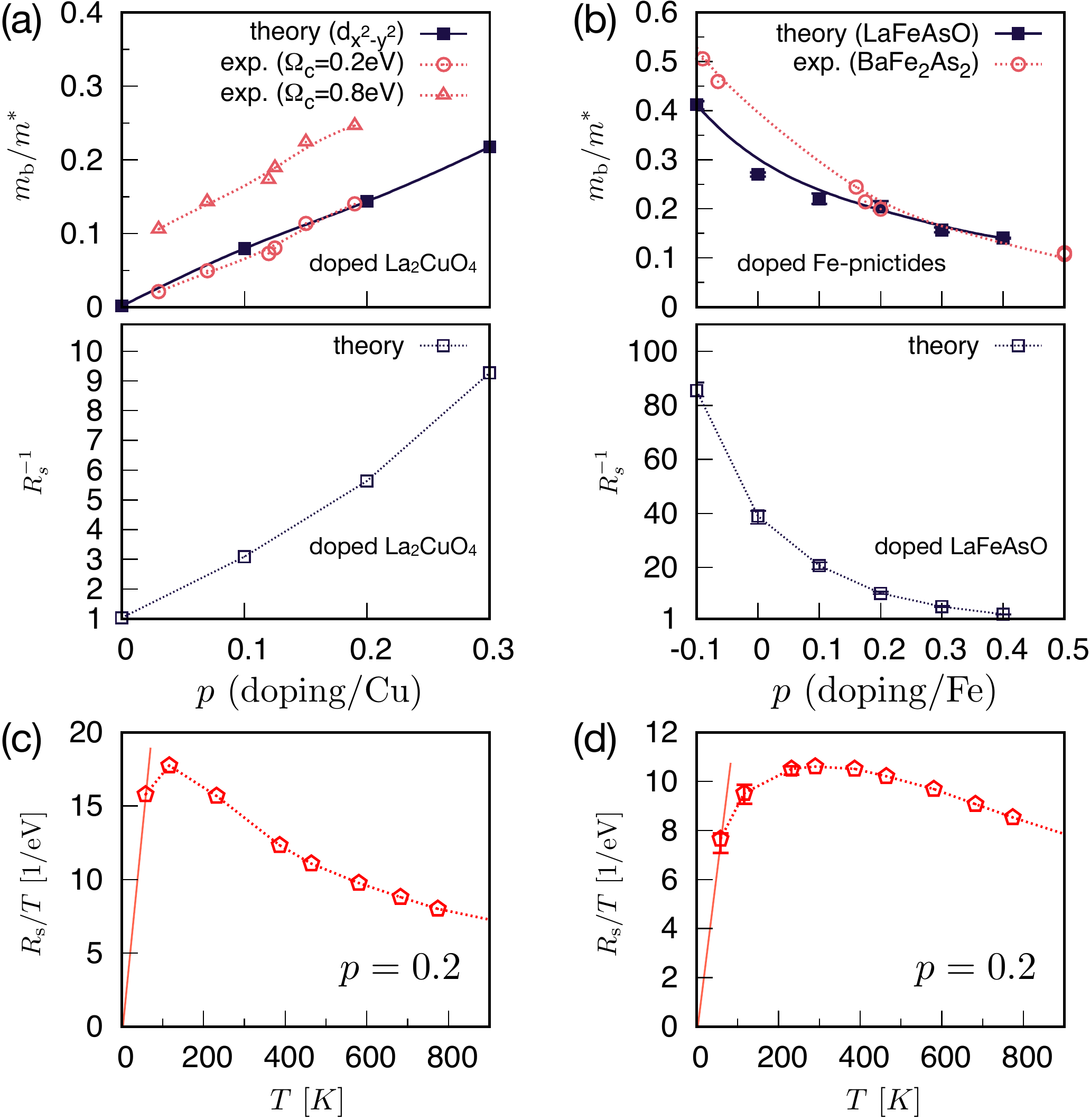}
	\caption{ (a--b) Doping dependence of the inverse of quasiparticle mass enhancement ($m_\mathrm{b}/m^*$) and $R_\mathrm{s}^{-1}$ in (a) La$_2$CuO$_4$ and (b) FeSCs. 
	Experimental estimates of $m_\mathrm{b}/m^*$ are taken from the Drude weights (obtained by integrating the optical conductivity data up to a cut-off frequency $\Omega_c$) \cite{Lucarelli,Ortolani,Millis} for (a), and the Sommerfeld coefficients \cite{Medici_2014,Hardy,Popovich,Pramanik,Mu,JSKim,Abdel-Hafiez,Storey} for (b). Theoretical estimates are obtained at $T=116$~K. (c--d) The calculated $T$ dependence of $R_\mathrm{s}/T$ for (c) 0.2-hole doped La$_2$CuO$_4$ and (d) 0.2-hole doped LaFeAsO.
	}
	\label{fig5}
\end{figure}

Let us first address the case of La$_2$CuO$_4$ which is a Mott insulator at $p=0$. Note first that our DMFT calculations using {\it ab initio} parameters correctly capture the Mott insulating phase at $p=0$, namely $m_\mathrm{b}/m^*$ of $d_{x^2-y^2}$ vanishes and $R_\mathrm{s}^{-1} \simeq 1$ [Fig.~\ref{fig4}(a)].
As $p$ increases (increasing hole doping), the correlation strength gets gradually reduced, which is also corroborated by the Drude weights of optical conductivity measurements \cite{Lucarelli,Ortolani,Millis} [Fig.~\ref{fig5}(a)]. This is the typical behavior of Mott systems.
Furthermore, the monotonic increment of both $m_\mathrm{b}/m^*$ and $R_\mathrm{s}^{-1}$ at least up to $p=0.3$ implies that Hund physics is not realized in La$_2$CuO$_4$ in this range of electron density.
This result is a direct consequence of a sizable $\Delta$ between the two $e_g$ orbitals ($\Delta/J \simeq 4.73$ \cite{Hirayama}), which suppresses significantly the Hund fluctuations by favoring large orbital polarization.

We now turn to the case of FeSCs which have been highlighted over many years as materials realization of Hund metal \cite{Georges}. 
While basically the same features are expected for the entire family, we take LaFeAsO for simplicity. The upper panel of Fig.~\ref{fig5}(b) presents the calculated $m_\mathrm{b}/m^*$  from the Sommerfeld coefficient ratio ($\gamma_\mathrm{b}/\gamma^*$) in comparison with the available experimental data for hole-doped BaFe$_2$As$_2$ \cite{Hardy,Popovich,Pramanik,Mu,JSKim,Abdel-Hafiez,Storey}.  
Here $\gamma_\mathrm{b}$ and $\gamma^*$ are band theory and DMFT (or specific heat) estimates, respectively. Specifically,  $m_\mathrm{b}/m^*=\gamma_\mathrm{b}/\gamma^*$ using a formula: $\gamma_\mathrm{b}/\gamma^*=\sum_{\eta}A_{\mathrm{b},\eta} / \sum_{\eta}A_{\eta}Z_\eta^{-1}$ where $A_{\mathrm{b},\eta}$ and $A_{\eta}$, respectively, are the density of states of orbital $\eta$ at the Fermi level obtained from the band theory and the DMFT \cite{Kotliar}. Note that the formula is strictly valid in a FL regime, so our calculated $\gamma_\mathrm{b}/\gamma^*$ should not be taken seriously as quantitative estimates.

Both theoretical and experimental $m_\mathrm{b}/m^*$ displayed in Fig.~\ref{fig5}(b) clearly demonstrate that the correlation strength increases with hole doping $p$ in FeSCs, which is in sharp contrast to the case of La$_2$CuO$_4$. The doping dependence of $R_\mathrm{s}^{-1}$ further supports the behavior.
Based on our scheme, this characteristic feature of $R_\mathrm{s}^{-1}$, namely $\partial R_\mathrm{s}^{-1} / \partial p<0$, corroborates that FeSCs are governed by strong Hund fluctuations.

Looking at the $T$ dependence of $R_\mathrm{s}$ [Figs.~\ref{fig5}(c--d)], both La$_2$CuO$_4$ and LaFeAsO at $0.2$-hole doping deviate clearly from the FL behavior ($R_\mathrm{s}/T \propto T$; solid lines are guides for the eye) down to $T \simeq 58$~K. Considering that the critical temperature $T_\mathrm{c}$ below which the superconductivity emerges is $T_\mathrm{c} \sim 40~\mathrm{K}$ in La$_{0.8}$Sr$_{0.2}$CuO$_4$ and in some FeSCs like Ba$_{0.6}$K$_{0.4}$Fe$_2$As$_2$ \cite{Rotter}, the non-FL behavior may be relevant for the emergence of the superconductivity at this doping. 

\section{Possible experimental probes}

We now discuss possible experiments to detect $R_\mathrm{s}$, which requires us to measure both short-time ($\tau=0$) and long-time ($\tau=1/2T$) local spin-spin correlation functions.
Although some subtleties exist, one can resort to x-ray emission or absorption spectroscopy and inelastic neutron scattering (INS). The x-ray techniques can measure the instantaneous local spin moments, $\mu_\mathrm{loc} = \sqrt{3C(0)}$, by probing the local fluctuations in the femtosecond scale as discussed in the context of FeSCs \cite{Gretarsson_2011,Gretarsson_2013,Pelliciari}. INS, on the other hand, can probe the long-time (or low-energy) fluctuations by effectively measuring the imaginary part of the dynamical spin susceptibility, $\mathrm{Im}\chi(\mathbf{k},\omega)$ ($\mathbf{k}$: crystal-momentum, $\omega$: real frequency), with the assumption that orbital moments are quenched \cite{Toschi,Chen_ferromagnet}. Using the following relation,
\begin{align}
C(\tau)  = \int d\mathbf{k} \int^{\infty}_{-\infty}  d\omega \frac{e^{-\omega \tau}}{1  - e^{-\frac{\omega}{T}}  } \mathrm{Im}\chi(\mathbf{k},\omega) ,
\label{eq6}
\end{align}
the long-time value $C(1/2T)$ can be approximated to
the local part of $\mathrm{Im}\chi(\mathbf{k},0)$, namely $\mathrm{Im}\chi_\mathrm{loc}(\omega=0)$. This is because $\mathrm{Im}\chi(\mathbf{k},0)$ is the dominant contribution to the $\omega$-integration when $\tau=1/2T$ due to a characteristic structure of the kernel in Eq.~(\ref{eq6}) \cite{Randeria}. 
The combination of these two spectroscopies hopefully provides a way to estimate $R_\mathrm{s}$. 

If we rely on the quasiparticle weight $Z$ for the experimental idenfication of Hund correlations instead of using $R_\mathrm{s}$, any probes that can measure the mass enhancement are relevant. As we have already noticed from Figs.~\ref{fig5}(a--b), the Drude weight of optical conductivity and the Sommerfeld coefficient of specific heat are standard techniques to estimate $m_\mathrm{b}/m^*$, provided $m_\mathrm{b}$ is given by the band theory. Furthermore, angle-resolved photoemission spectroscopy (ARPES) enables us to extract the orbital dependent contributions, $(m_\mathrm{b}/m^*)_\eta$. Assisted by these experimental techniques, it is feasible to estimate the sign of $\partial Z / \partial p$.

\section{Discussion}

We finally remark on the related open questions. While our approach based on the sign of $\partial R_\mathrm{s}^{-1}/\partial p$ should be valid in most of the transition-metal compounds to which the Kanamori (or Slater) type of local interaction [Eq.~(\ref{eq1})] is relevant, its applicability to materials with more complicated interactions remains to be resolved. It is also worth investigating the effect of interorbital hopping which plays a crucial role in Mott metal-to-insulator transitions in non-degenerate-orbital systems \cite{Kugler_OSMP,Stepanov_OSMP,Ryee_moire}.
Nonlocal electron correlations and possible symmetry-breaking transitions affect the low-temperature physics and thus may influence our picture. We expect, however, that measuring $R_\mathrm{s}$ should not be hindered by the latter if the measurement is done above the transition temperature. Further studies are required to clarify these issues.

\section{Acknowledgments}
S.R. is grateful to Tim Wehling for many stimulating conversations. S.R. and M.J.H. were supported by the National Research Foundation of Korea (Grant Nos.~2021R1A2C1009303 and NRF-2018M3D1A1058754). S.R. was also supported by the KAIX Fellowship. S.C. was supported by the U.S. Department of Energy, Office of Science, Basic Energy Sciences as a part of the Computational Materials Science Program. S.C. was also supported by a KIAS individual Grant (No. CG090601) at Korea Institute for Advanced Study. This research used resources of the National Energy Research Scientific Computing Center (NERSC), a U.S. Department of Energy Office of Science User Facility operated under Contract No. DE-AC02-05CH11231.

\appendix

\section{$Z$ and $R_\mathrm{s}^{-1}$ for a two-degenerate-orbital model on a Bethe lattice} \label{appendix_A}

\begin{figure*} [!htbp]
	\includegraphics[width=0.9\textwidth, angle=0]{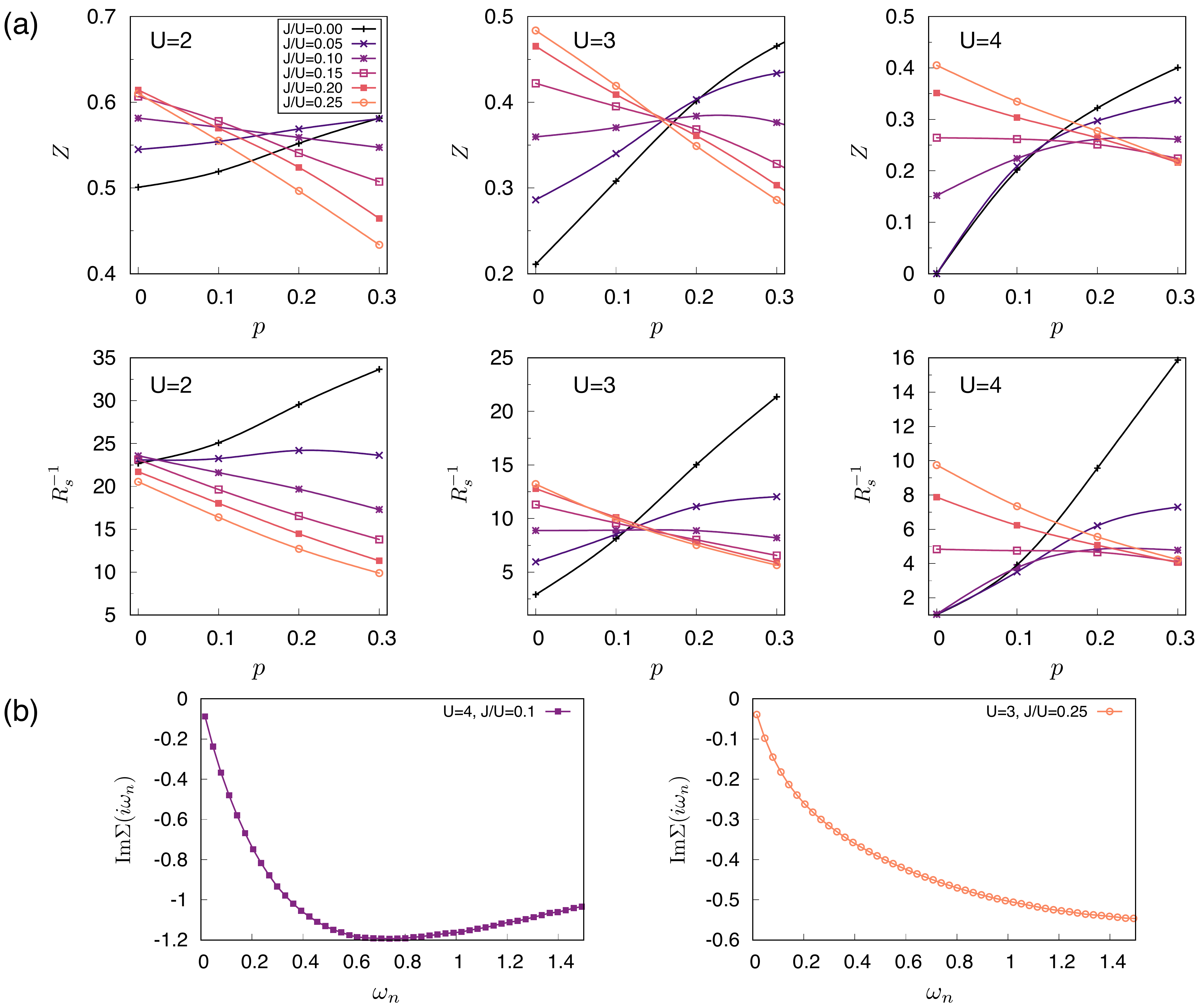}
	\caption{(a) $Z$ and $R_\mathrm{s}^{-1}$ as a function of $p$ for a two-degenerate-orbital model on a Bethe lattice at three different values of $U$ ($U=2$, 3, and 4). The electron filling $n$ corresponds to $n=3-p$ for all the cases. We used a higher simulation temperature for the calculation of $R_\mathrm{s}^{-1}$  ($T=0.02$) than for $Z$ ($T=0.005$) so that $ \langle S_z(\tau) S_z(0) \rangle \rvert_{\tau=1/(2T)} \gg 0$. (b) The self-energy on the imaginary frequency axis $\mathrm{Im}\Sigma(i\omega_n)$ obtained from DMFT. Left: $U=4$ and $J/U=0.1$ at $p=0$. Right: $U=3$ and $J/U=0.25$ at $p=0.3$.}
	\label{fig6}
\end{figure*}

Figure~\ref{fig6} displays additional data for $Z$ vs.~$p$ and $R_\mathrm{s}^{-1}$ vs.~$p$ at three different values of $U$. 
Here, $Z = \big(1-{\partial \mathrm{Im}\Sigma(i\omega_n)}/{\partial \omega_n} \big|_{\omega_n \to 0+}\big)^{-1}$ where $\Sigma(i\omega_n)$ is the local self-energy on the imaginary frequency axis; see, e.g., Fig.~\ref{fig6}(b). We fitted a fourth-order polynomial to the self-energies in the lowest six imaginary frequency points, following Refs.~\cite{Mravlje,Ryee3}.

One can notice from Fig.~\ref{fig6} that the value of $J/U$ above which $\partial Z/ \partial p <0$ and $\partial R_\mathrm{s}^{-1}/ \partial p <0$ increases as $U$ is increased; the Mott behavior becomes predominant even up to a fairly large value of $J/U$. Thus, the effect of $J$ gets weakened as $U$ is increased.
A rationale behind this phenomenon can be drawn from the generic form of Kondo couplings obtained via the Schrieffer-Wolff transformation of multiorbital impurity models, which read $\mathcal{J}_{i}/V^2 = \mathcal{J}^{J=0}_i/V^2 + \mathcal{O}({J}/{U^2}) + \mathcal{O}({J^2}/{U^3}) + \cdots$ [Eqs.~(\ref{eqS13})--(\ref{eqS15})]. Thus, at a regime of $J \ll U$, $\mathcal{J}_i \simeq \mathcal{J}^{J=0}_i$ by which the effect of Hund coupling $J$ on $\mathcal{J}_i$s becomes largely suppressed.

\section{$Z$ and $R_\mathrm{s}^{-1}$ for a two-degenerate-orbital model on a square lattice} \label{appendix_B}

\begin{figure*} [!htbp]
	\includegraphics[width=0.9\textwidth, angle=0]{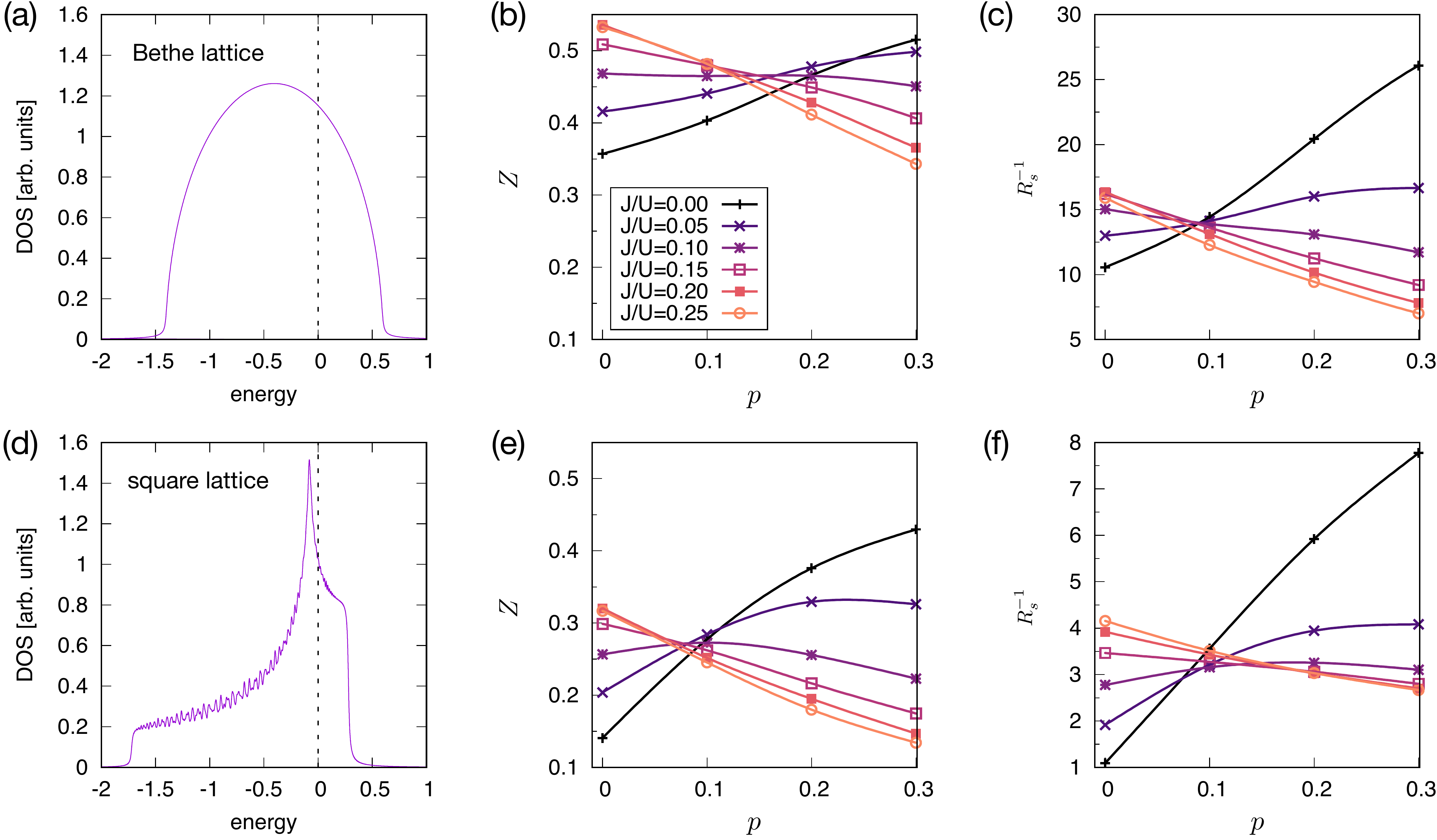}
	\caption{The density of states, $Z$ vs.~$p$, and $R_\mathrm{s}^{-1}$ vs.~$p$ for (a--c) an infinite-dimensional Bethe lattice and (d--f) a two-dimensional square lattice with NN and NNN hopping amplitudes, $t=0.25$ and $t'=0.08$, respectively, for degenerate two orbitals. $D=1$, $U=2.5$, and $n=3-p$ for both lattices. The vertical dashed lines in (a) and (d) denote the Fermi level for $p=0$. We used a higher simulation temperature for the calculations of $R_\mathrm{s}^{-1}$  ($T=0.02$) than for $Z$ ($T=0.005$) so that $ \langle S_z(\tau) S_z(0) \rangle \rvert_{\tau=1/(2T)} \gg 0$.}
	\label{fig7}
\end{figure*}

In the main text, we mainly focus on an infinite-dimensional Bethe lattice with semielliptical density of states (DOS) in order to focus on generic features rather than material specific ones. In realistic lattices such as a square lattice, a van Hove singularity (vHS) can exist near the Fermi level. This singularity features a divergence in the DOS [Fig.~\ref{fig7}(d)], largely affecting the strength of electron correlations by effectively suppressing low-energy hopping processes; see, e.g., Refs~\cite{Mravlje,Karp,HJLee2,HJLee} for related discussions. 
Thus, one may ask whether this vHS has any effects on the signs of $\partial Z/ \partial p$ and $\partial R_\mathrm{s}^{-1}/ \partial p$.

Figure~\ref{fig7} presents $Z$ and $R_\mathrm{s}^{-1}$ as a function of $p$ for a square lattice with nearest-neighbor (NN) and next-nearest-neighbor (NNN) hopping amplitudes, $t=0.25$ and $t'=0.08$, respectively. Namely, the kinetic part of our Hamiltonian for the square lattice reads $H_\mathrm{K} = -t\sum_{\langle ij \rangle,\sigma} d^{\dagger}_{i\sigma}d_{j\sigma} - t'\sum_{\langle \langle ij \rangle \rangle,\sigma} d^{\dagger}_{i\sigma}d_{j\sigma}$, where $\langle ij \rangle$ and $\langle \langle ij \rangle \rangle$ denotes, respectively, the NN and NNN sites.
Here, $D=1$ and $U=2.5$. The results of the Bethe lattice are also plotted for comparison. While the correlation strength itself is enhanced in the square lattice rather than in the Bethe lattice due to the presence of the vHS near the Fermi level, qualitatively the similar behavior is observed for both $Z$ and $R_\mathrm{s}^{-1}$: $\partial Z/ \partial p$ and $\partial R_\mathrm{s}^{-1}/ \partial p$ change their signs from plus to minus by $J/U$ or by $p$.

\section{$R_\mathrm{s}^{-1}$ for a three-degenerate-orbital model on a Bethe lattice}
\label{appendix_C}

\begin{figure*} [!htbp]
	\includegraphics[width=0.9\textwidth, angle=0]{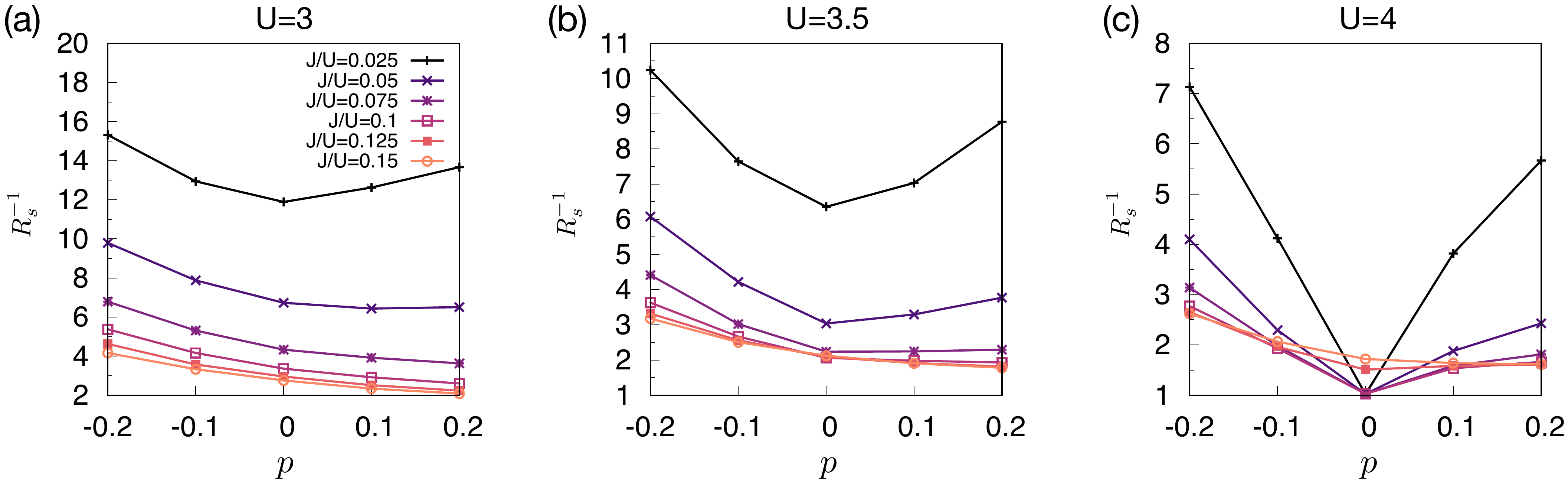}
	\caption{$R_\mathrm{s}^{-1}$ as a function of hole doping $p$ at $T=0.02$ for a three-degenerate-orbital model on a Bethe lattice at (a) $U=3$, (b) $U=3.5$, and (c) $U=4$. $n=4-p$.}
	\label{fig8}
\end{figure*}

A three-degenerate-orbital model has served as a prototypical system for Hund metal physics \cite{Janus,Georges}. Here, unlike the cases of two-degenerate-orbital models, $J$ lifts degeneracy of the ground state atomic multiplets even when $p=0$, whereby $J$ strongly enhances the correlation strength by forming a large composite spin moment \cite{Nevidomskyy,Janus,Georges}. As can be clearly seen from Fig.~\ref{fig8}, the overall shape of $R_\mathrm{s}^{-1}$ gradually changes from ``V-shape" to a monotonic behavior as $J$ increases.

At any rate, for the $p>0$ side, $\partial R_\mathrm{s}^{-1}/\partial p$ changes its sign by $J$ as is discussed for two-orbital models; see Fig.~\ref{fig8}. Note here that very low-$T$ calculations are required to reach the coherence temperature below which the long-lived quasiparticles are formed in three-orbital models \cite{Georges,Kowalski}, which is computationally demanding for our DMFT calculations adopting a continuous-time quantum Monte Carlo algorithm. Hence, $Z$ may not be a good measure of the correlation strength even for the lowest temperature practically accessible within our computation scheme. We thus present only $R_\mathrm{s}^{-1}$ in Fig.~\ref{fig8}.

\section{Kondo couplings from the Schrieffer-Wolff transformation} \label{appendix_D}

Here, we derive Kondo couplings by applying the canonical SW transformation to a relevant impurity model. We follow the strategy depicted in Refs.~\cite{Yin_power,SOS1,SOS2}. 

Note first that the original lattice model is mapped onto an auxiliary impurity model in DMFT. Let us thus write down the impurity Hamiltonian with $\mathrm{SU}(2)_\mathrm{spin} \otimes \mathrm{SU}(M)_\mathrm{orbital}$ symmetry:
\begin{align}
	H_\mathrm{imp} = H_\mathrm{loc} + H_\mathrm{bath}  +  H_\mathrm{hyb},
	\label{eqS1}
\end{align}
where
\begin{align}
	H_\mathrm{loc} &=  U\sum_{\eta}{n_{\eta \uparrow} n_{\eta \downarrow}} 
	+ \sum_{\eta < \eta',\sigma\sigma'}(U'-J\delta_{\sigma\sigma'}){ n_{ \eta \sigma} n_{ \eta' \sigma'}} 
    \nonumber \\ &+ J\sum_{\eta \neq \eta'}d^{\dagger}_{\eta \uparrow}  d^{\dagger}_{\eta' \downarrow} d_{\eta \downarrow} d_{\eta' \uparrow}  +\sum_{\eta,\sigma}(\epsilon_\eta - \mu)n_{\eta\sigma},  \label{eqS2}\\
	H_\mathrm{bath} &= \sum_{\mathbf{k}\eta\sigma}\epsilon_{\mathbf{k}\eta} \psi^{\dagger}_{\mathbf{k}\eta\sigma}\psi_{\mathbf{k}\eta\sigma}, \\
	H_\mathrm{hyb} &= \sum_{\eta\sigma}V \psi^{\dagger}_{\eta\sigma}d_{\eta\sigma} + \mathrm{h.c.},
\end{align}
Here, $\psi^{\dagger}$ ($\psi$) is the creation (destruction) operator for bath states. $U'=U-J$. We will later get back to the form of Eq.~(\ref{eq1}) which includes the pair-hopping term and $U'=U-2J$ for $H_\mathrm{loc}$. 

Our goal is, by integrating out valence fluctuations, to construct an effective Kondo model for low-energy physics:
\begin{align}
	H_\mathrm{eff} = H_\mathrm{int} +  H_\mathrm{bath},
\end{align}
where $H_\mathrm{int}$ is given by the SW transformation:
\begin{widetext}
\begin{equation}
	    \begin{split}
		H_\mathrm{int} = &-P^{n_0} H_\mathrm{hyb} \Bigg\{ \sum_{i}\frac{P^{n_0+1}_i}{E^{n_0+1}_i} + \sum_{i}\frac{P^{n_0-1}_i}{ E^{n_0-1}_i} \Bigg\} H_\mathrm{hyb} P^{n_0}  \\
		= &-\sum_{i,\{ \eta \}, \{ \sigma\} } \frac{V^2}{E^{n_0+1}_i}\psi^{\dagger}_{\eta_1\sigma_1} \psi_{\eta_2\sigma_2} P^{n_0} d_{\eta_3\sigma_3} P^{n_0+1}_i d^{\dagger}_{\eta_4\sigma_4} P^{n_0}  \delta_{\eta_1 \eta_3} \delta_{\eta_2 \eta_4} \delta_{\sigma_1 \sigma_3} \delta_{\sigma_2 \sigma_4}   \\ 
		& -  \sum_{i,\{ \eta \}, \{ \sigma \} } \frac{V^2}{E^{n_0-1}_i}\psi_{\eta_1\sigma_1} \psi^{\dagger}_{\eta_2\sigma_2} P^{n_0} d^{\dagger}_{\eta_3\sigma_3} P^{n_0-1}_i d_{\eta_4\sigma_4} P^{n_0} \delta_{\eta_1 \eta_3} \delta_{\eta_2 \eta_4} \delta_{\sigma_1 \sigma_3} \delta_{\sigma_2 \sigma_4},
		\label{eqS6} 
		\end{split}
\end{equation}
\end{widetext}
with $P^{n_0}$ being the projector to the atomic ground state multiplet (with eigenvalue $\varepsilon^{n_0}$) in charge $N=n_0$ subspace of Eq.~(\ref{eqS2}). $P^{n_0\pm1}_i$ project onto the atomic multiplets of Eq.~(\ref{eqS2}) having eigenvalues of $\varepsilon^{n_0\pm1}_i$ in charge $n_0\pm1$ subspaces. The subscript $i$ is the index for labeling different multiplets. $E^{n_0\pm1}_i \equiv \varepsilon^{n_0\pm1}_i - \varepsilon^{n_0} $ refers to the charge excitation energy.

Now, we use the following relation which holds for $\mathrm{SU}(M)$ symmetry:
\begin{align}
	\delta_{il}\delta_{kj} = \frac{1}{M} \delta_{ij}\delta_{kl} + \frac{1}{2}\sum_{\alpha}\tau^\alpha_{ij}\tau^\alpha_{kl},
	\label{eqS7}
\end{align}
where $\tau$ is the generator of the $\mathrm{SU}(M)$ symmetric group, namely, $\tau^\alpha$ corresponds to Pauli matrices ($\sigma^\alpha$) for $M=2$, and to Gell-Mann matrices for $M=3$. We hereafter use the Einstein summation convention for simplicity. Inserting Eq.~(\ref{eqS7}) into Eq.~(\ref{eqS6}) for both spin and orbital leads to the following form of $H_\mathrm{int}$:
\begin{equation}
	\begin{split}
	H_\mathrm{int} &= \mathcal{J}_\mathrm{p} \psi^{\dagger}_{\eta\sigma}\psi_{\eta\sigma} \\ & + \mathcal{J}_s S^\alpha \psi^{\dagger}_{\eta\sigma} \Big( \frac{\sigma^\alpha_{\sigma \sigma'}}{2} \Big) \psi_{\eta\sigma'} \\ &+ \mathcal{J}_\mathrm{o} T^\alpha \psi^{\dagger}_{\eta\sigma} \Big( \frac{\tau^\alpha_{\eta \eta'}}{2} \Big) \psi_{\eta'\sigma} \\ &+ \mathcal{J}_\mathrm{so} S^\alpha T^\beta \psi^{\dagger}_{\eta\sigma} \Big( \frac{ \sigma^\alpha_{\sigma \sigma'}}{2} \frac{\tau^\beta_{\eta \eta'}}{2} \Big) \psi_{\eta'\sigma'} \label{eqS8}
	\end{split}
\end{equation}
$S^\alpha = d^{\dagger}_{\eta \sigma} (\sigma^\alpha_{\sigma \sigma'}/2) d_{\eta \sigma'}$ and $T^\beta = d^{\dagger}_{\eta \sigma} (\tau^\beta_{\eta \eta'}/2) d_{\eta' \sigma}$ are the spin and orbital operators for impurity degrees of freedom. $\mathcal{J}_\mathrm{p}$, $\mathcal{J}_\mathrm{s}$, $\mathcal{J}_\mathrm{o}$, and $\mathcal{J}_\mathrm{so}$ are Kondo couplings for potential scattering, spin, orbital, and spin-orbital terms, which are given by
\begin{widetext}
\begin{align}
	\mathcal{J}_\mathrm{p} \langle \phi_0 | I_S \otimes I_T | \phi_0 \rangle &= -\frac{V^2}{2M}\Bigg\{ \frac{\langle \phi_0 | d_{\eta \sigma} P^{n_0+1}_i d^{\dagger}_{\eta \sigma} | \phi_0 \rangle}{E^{n_0+1}_i} - \frac{\langle \phi_0 | d^{\dagger}_{\eta \sigma} P^{n_0-1}_i d_{\eta \sigma} | \phi_0 \rangle}{E^{n_0-1}_i} \Bigg\}, \\
	\mathcal{J}_\mathrm{s} \langle \phi_0 | S^{\alpha} \otimes I_T | \phi_0 \rangle &= -\frac{V^2}{M} \sigma^{\alpha}_{\sigma \sigma'} \Bigg\{ \frac{\langle \phi_0 | d_{\eta \sigma'} P^{n_0+1}_i d^{\dagger}_{\eta \sigma} | \phi_0 \rangle}{E^{n_0+1}_i} - \frac{\langle \phi_0 | d^{\dagger}_{\eta \sigma} P^{n_0-1}_i d_{\eta \sigma'} | \phi_0 \rangle}{E^{n_0-1}_i} \Bigg\}, \label{eqS10}\\
	\mathcal{J}_\mathrm{o} \langle \phi_0 | I_S \otimes T^\beta | \phi_0 \rangle &= -\frac{V^2}{2} \tau^{\beta}_{\eta \eta'} \Bigg\{ \frac{\langle \phi_0 | d_{\eta' \sigma} P^{n_0+1}_i d^{\dagger}_{\eta \sigma} | \phi_0 \rangle}{E^{n_0+1}_i} - \frac{\langle \phi_0 | d^{\dagger}_{\eta \sigma} P^{n_0-1}_i d_{\eta' \sigma} | \phi_0 \rangle}{E^{n_0-1}_i} \Bigg\}, \label{eqS11}\\
	\mathcal{J}_\mathrm{so} \langle \phi_0 | S^{\alpha} \otimes T^\beta | \phi_0 \rangle &= -V^2 \sigma^{\alpha}_{\sigma \sigma'} \tau^{\beta}_{\eta \eta'} \Bigg\{ \frac{\langle \phi_0 | d_{\eta' \sigma'} P^{n_0+1}_i d^{\dagger}_{\eta \sigma} | \phi_0 \rangle}{E^{n_0+1}_i} - \frac{\langle \phi_0 | d^{\dagger}_{\eta \sigma} P^{n_0-1}_i d_{\eta' \sigma'} | \phi_0 \rangle}{E^{n_0-1}_i} \Bigg\}. \label{eqS12}
\end{align}
\end{widetext}
Here, $| \phi_0 \rangle$ denotes the atomic ground state multiplet in the charge $n$ subspace. Since the first term in Eq.~(\ref{eqS8}) is irrelevant for dynamics of local moments \cite{SOS1,SOS2}, we discard $\mathcal{J}_\mathrm{p}$ from our discussion.


\begin{table*} [!htbp] 
	\renewcommand{\arraystretch}{1.5}
	\begin{tabular}{c  c  c  c  c  c  }
		\hline \hline
		Index &\ \ Eigenstate &\ \  $N$ &\ \ $S$ &\ \ $S_z$ &\ \ Eigenvalue \\
		\hline \hline			
		1 & \ \ $|0,0\rangle$   &\ \ 0  &\ \ 0 &\ \  0 &\ \  0  \\
		\hline
		2 & \ \ $|0,\uparrow\rangle$&\ \     1 &\ \ 1/2 & \ \ 1/2 &\ \  $-\mu$ \\
		3 & \ \ $|\uparrow,0\rangle$&\ \     1 &\ \ 1/2 & \ \ 1/2 &\ \  $-\mu$ \\
		4 & \ \ $|0,\downarrow\rangle$&\ \     1 &\ \ 1/2 & \ \ -1/2 &\ \  $-\mu$ \\
		5 & \ \ $|\downarrow,0\rangle$&\ \    1 &\ \ 1/2 & \ \ -1/2 &\ \  $-\mu$ \\		
		\hline
		6 & \ \ $|\uparrow,\uparrow\rangle$ &\ \  2 &\ \ 1 &\ \ 1 &\ \  $U-2J-2\mu$ \\ 
		7 & \ \ $\big(|\uparrow,\downarrow\rangle + |\downarrow,\uparrow\rangle\big)/\sqrt{2}$ &\ \  2 &\ \ 1 &\ \ 0 &\ \  $U-2J-2\mu$ \\ 
		8 & \ \ $|\downarrow,\downarrow\rangle$ &\ \   2 &\ \ 1 &\ \ -1 &\ \  $U-2J-2\mu$ \\
		9 & \ \ $|\uparrow \downarrow,0\rangle$ &\ \  2 &\ \ 0 &\ \ 0 &\ \  $U-2\mu$ \\ 
		10 & \ \ $\big(|\uparrow,\downarrow \rangle - |\downarrow,\uparrow\rangle\big)/\sqrt{2}$ &\ \  2 &\ \ 0 &\ \ 0 &\ \  $U-2\mu$ \\ 
		11 & \ \ $|0,\uparrow \downarrow \rangle$ &\ \  2 &\ \ 0 &\ \ 0 &\ \  $U-2\mu$ \\ 
		\hline
		12 & \ \ $|\uparrow\downarrow,\uparrow\rangle$&\ \  3 &\ \ 1/2 & \ \ 1/2 &\ \  $3U-3J-3\mu$ \\
		13 & \ \ $|\uparrow,\uparrow\downarrow\rangle$&\ \  3 &\ \ 1/2 & \ \ 1/2 &\ \  $3U-3J-3\mu$ \\
		14 & \ \ $|\uparrow\downarrow,\downarrow\rangle$&\ \  3 &\ \ 1/2 & \ \ -1/2 &\ \  $3U-3J-3\mu$ \\
		15 & \ \ $|\downarrow,\uparrow\downarrow\rangle$&\ \ 3 &\ \ 1/2 & \ \ -1/2 &\ \  $3U-3J-3\mu$ \\		
		\hline	   
		16 & \ \ $|\uparrow \downarrow, \uparrow \downarrow \rangle$&\ \ 4 &\ \ 0 &\ \ 0 &\ \ $6U-6J-4\mu$ \\
		\hline \hline
		
	\end{tabular}
	\caption{Eigenstates and eigenvalues of Eq.~(\ref{eqS2}) with $\epsilon_1=\epsilon_2=0$ obeying $\mathrm{SU(2)}_\mathrm{spin} \otimes \mathrm{SU(2)}_\mathrm{orbital}$ symmetry. The first entry in a ket of an eigenstate is a state of orbital-1 and the second is of orbital-2. }
	\label{table_s1}
\end{table*}

While the discussion below is valid for any $\mathrm{SU}(2) \otimes \mathrm{SU}(M)$ models, let us focus on the case of two orbitals ($M=2$) with $n_0=3$. For generic $\mathrm{SU}(2) \otimes \mathrm{SU}(M)$ cases, refer to Ref.~\cite{SOS1}. Eigenstates and eigenvalues of Eq.~(\ref{eqS2}) are listed in Table~\ref{table_s1}.
We have the freedom to choose $| \phi_0 \rangle$, $\alpha$, and $\beta$ to evaluate Eqs.~(\ref{eqS10})--(\ref{eqS12}). Hence, for convenience, $| \phi_0 \rangle = |\uparrow,\uparrow\downarrow\rangle$ and $\alpha=\beta=3$. Kondo couplings are now given by:
\begin{widetext}
\begin{align}
	\mathcal{J}_\mathrm{s} &= \frac{V^2}{2} \Bigg( \frac{2}{E^{n_0+1}_{16}}  -  \frac{2}{E^{n_0-1}_{6}} +  \frac{1}{E^{n_0-1}_{7}} +  \frac{1}{E^{n_0-1}_{10}} +  \frac{2}{E^{n_0-1}_{11}}   \Bigg)  = \frac{V^2}{2} \Bigg( \frac{2}{ E_0^{|4,0\rangle}} -\frac{1}{ E_0^{|2,1\rangle}} + \frac{3}{ E_0^{|2,0\rangle}}  \Bigg), \label{eqS13}\\
	\mathcal{J}_\mathrm{o} &= \frac{V^2}{2} \Bigg( \frac{2}{E^{n_0+1}_{16}}  +  \frac{2}{E^{n_0-1}_{6}} +  \frac{1}{E^{n_0-1}_{7}} +  \frac{1}{E^{n_0-1}_{10}} -  \frac{2}{E^{n_0-1}_{11}}   \Bigg)  = \frac{V^2}{2} \Bigg( \frac{2}{ E_0^{|4,0\rangle}} +\frac{3}{ E_0^{|2,1\rangle}} - \frac{1}{ E_0^{|2,0\rangle}}  \Bigg), \label{eqS14}\\
	\mathcal{J}_\mathrm{so} &= 2V^2 \Bigg( \frac{2}{E^{n_0+1}_{16}} +  \frac{2}{E^{n_0-1}_{6}} -  \frac{1}{E^{n_0-1}_{7}} -  \frac{1}{E^{n_0-1}_{10}}  +  \frac{2}{E^{n_0-1}_{11}}   \Bigg)  = 2V^2 \Bigg( \frac{2}{ E_0^{|4,0\rangle}} +\frac{1}{ E_0^{|2,1\rangle}} + \frac{1}{ E_0^{|2,0\rangle}} \Bigg), \label{eqS15}
\end{align}
\end{widetext}
where the subscript $i$ of $E^{n_0\pm1}_i$ ($\equiv \varepsilon^{n_0\pm1}_i - \varepsilon^{n_0}_{13}$) refers to the index of the eigenstate in Table~\ref{table_s1}. In the second equalities of the above equations, we re-write terms using $E_k^{|N,S\rangle}$ which denotes the excitation energy from the ground state $|3,1/2\rangle $ to the excited atomic multiplet $|N,S \rangle$. The subscript $k$ ($k \in \{0,1,...\}$) refers to the $k$-th lowest eigenvalue in the corresponding $|N,S\rangle$ subspace. We henceforth restrict ourselves to a region where $E_k^{|N,S\rangle}>0$ and $p>0$.

\begin{table*} [!htbp] 
	\renewcommand{\arraystretch}{1.5}
	\begin{tabular}{c  c  c  c  c  c }
		\hline \hline
		Index &\ \  Eigenstate &\ \ $N$ &\ \ $S$ &\ \ $S_z$ &\ \ Eigenvalue \\
		\hline \hline	
		
		1 &\ \  $|0,0\rangle$ &\ \ 0  &\ \ 0 &\ \  0 &\ \  0  \\
		\hline
		2 &\ \  $|0,\uparrow\rangle$&\ \ 1 &\ \ 1/2 & \ \ 1/2 &\ \  $-\mu$ \\
		3 &\ \  $|\uparrow,0\rangle$&\ \ 1 &\ \ 1/2 & \ \ 1/2 &\ \  $-\mu$ \\
		4 &\ \  $|0,\downarrow\rangle$&\ \ 1 &\ \ 1/2 & \ \ -1/2 &\ \  $-\mu$ \\
		5 &\ \  $|\downarrow,0\rangle$&\ \ 1 &\ \ 1/2 & \ \ -1/2 &\ \  $-\mu$ \\		
		\hline
		6 &\ \  $|\uparrow,\uparrow\rangle$ &\ \ 2 &\ \ 1 &\ \ 1 &\ \  $U-3J-2\mu$ \\ 
		7 &\ \  $\big(|\uparrow,\downarrow\rangle + |\downarrow,\uparrow\rangle\big)/\sqrt{2}$ &\ \ 2 &\ \ 1 &\ \ 0 &\ \  $U-3J-2\mu$ \\ 
		8 &\ \  $|\downarrow,\downarrow\rangle$ &\ \ 2 &\ \ 1 &\ \ -1 &\ \  $U-3J-2\mu$ \\
		9 &\ \  $\big(|\uparrow,\downarrow\rangle - |\downarrow,\uparrow\rangle\big)/\sqrt{2}$ &\ \ 2 &\ \ 0 &\ \ 0 &\ \  $U-J-2\mu$ \\ 
		10 &\ \  $\big(|\uparrow\downarrow,0\rangle - |0,\downarrow\uparrow\rangle\big)/\sqrt{2}$ &\ \ 2 &\ \ 0 &\ \ 0 &\ \  $U-J-2\mu$ \\ 
		11 &\ \  $\big(|\uparrow\downarrow,0\rangle + |0,\downarrow\uparrow\rangle\big)/\sqrt{2}$ &\ \ 2 &\ \ 0 &\ \ 0 &\ \  $U+J-2\mu$ \\ 
		\hline
		12 &\ \  $|\uparrow\downarrow,\uparrow\rangle$&\ \ 3 &\ \ 1/2 & \ \ 1/2 &\ \  $3U-5J-3\mu$ \\
		13 &\ \  $|\uparrow,\uparrow\downarrow\rangle$&\ \ 3 &\ \ 1/2 & \ \ 1/2 &\ \  $3U-5J-3\mu$ \\
		14 &\ \  $|\uparrow\downarrow,\downarrow\rangle$&\ \ 3 &\ \ 1/2 & \ \ -1/2 &\ \  $3U-5J-3\mu$ \\
		15 &\ \  $|\downarrow,\uparrow\downarrow\rangle$&\ \ 3 &\ \ 1/2 & \ \ -1/2 &\ \  $3U-5J-3\mu$ \\		
		\hline	   
		16 &\ \  $|\uparrow \downarrow, \uparrow \downarrow \rangle$ &\ \ 4 &\ \ 0 &\ \ 0 &\ \ $6U-10J-4\mu$ \\
		\hline \hline
		
	\end{tabular}
	\caption{Eigenstates and eigenvalues of Eq.~(1) with $U'=U-2J$ and $\epsilon_1=\epsilon_2=0$. The first entry in a ket of an eigenstate is a state of orbital-1 and the second is of orbital-2.}
	\label{table_s2}
\end{table*}

We now consider responses of these couplings due to changes in filling. To mimic the effect of a small increase in $p$, let us consider a situation where $\mu$ is slightly decreased by $d\mu$ ($>0$), i.e., $\mu \rightarrow \mu- d\mu$. The concomitant changes in $\mathcal{J}_i$s are given by:
\begin{align}
	-&\Big( \frac{\partial \mathcal{J}_\mathrm{s}}{\partial \mu} \Big)  d\mu \nonumber  \\ &= \frac{V^2}{2} \Bigg\{ -\frac{2}{ \big(E^{|4,0\rangle}_0\big)^2 } -\frac{1}{ \big(E^{|2,1\rangle}_0\big)^2 } + \frac{3}{ \big(E^{|2,0\rangle}_0\big)^2 } \Bigg\} d\mu
	 \label{eqS16}\\
	-&\Big( \frac{\partial \mathcal{J}_\mathrm{o}}{\partial \mu} \Big)  d\mu  \nonumber \\ &= \frac{V^2}{2} \Bigg\{ -\frac{2}{ \big(E^{|4,0\rangle}_0\big)^2 } +\frac{3}{ \big(E^{|2,1\rangle}_0\big)^2 } - \frac{1}{ \big(E^{|2,0\rangle}_0\big)^2 } \Bigg\} d\mu
	 \label{eqS17}\\
	-&\Big( \frac{\partial \mathcal{J}_\mathrm{so}}{\partial \mu} \Big) d\mu \nonumber \\ &= 2V^2 \Bigg\{ -\frac{2}{ \big(E^{|4,0\rangle}_0\big)^2 } +\frac{1}{ \big(E^{|2,1\rangle}_0\big)^2 } + \frac{1}{ \big(E^{|2,0\rangle}_0\big)^2 } \Bigg\} d\mu
	 \label{eqS18}
\end{align}
When $J=0$, $E_0^{|2,1\rangle}$ is equal to $E_0^{|2,0\rangle}$. Since $E^{|4,0\rangle}_0$ is larger than $E_0^{|2,1\rangle}$ for the hole doped side, Eqs.~(\ref{eqS16})--(\ref{eqS18}) are all positive implying that all the Kondo coupling constants evolve in a way to weaken the correlation strength. When $J>0$ and $p>0$, on the other hand, $E_0^{|2,1\rangle}$ is smaller than the other $E_i^{|N,S\rangle}$s. In this case, Eqs.~(\ref{eqS16})--(\ref{eqS18}) are controlled mainly by the terms related to $E_0^{|2,1\rangle}$. Thus, we arrive at the following relations for $J>0$:
\begin{align}
	-\Big( \frac{\partial \mathcal{J}_\mathrm{s}}{\partial \mu} \Big) d\mu &\approx -\frac{1}{2}\frac{V^2}{\big(E^{|2,1\rangle}_0\big)^2 } d\mu
	\quad \mathrm{for}~\mathrm{spin}, \label{eqS19}\\
	-\Big( \frac{\partial \mathcal{J}_\mathrm{o}}{\partial \mu} \Big) d\mu &\approx  \frac{3}{2}\frac{V^2}{ \big(E^{|2,1\rangle}_0\big)^2 } d\mu
	\quad \mathrm{for}~\mathrm{orbital}, \\
	-\Big( \frac{\partial \mathcal{J}_\mathrm{so}}{\partial \mu} \Big) d\mu &\approx  2\frac{V^2}{ \big(E^{|2,1\rangle}_0\big)^2 } d\mu
	\quad \mathrm{for}~\mathrm{spin{\text -}orbital}. 
\end{align}
The above relations indicate that only $\mathcal{J}_\mathrm{s}$ decreases due to Hund fluctuations as $p$ is increased. In contrast, $\mathcal{J}_\mathrm{o}$ increases with $p$, favoring the screening of orbital degrees of freedom, which is consistent with the enhanced spin-orbital separation by $p$ for a finite $J$ in three-orbital models \cite{SOS2,Stadler2}.

\begin{table*} [!htbp]
	\renewcommand{\arraystretch}{1.5}
	\begin{tabular}{c  c  c  c  c  c }
		\hline \hline
		Index &\ \ Eigenstate &\ \ $N$ &\ \ $S$ &\ \ $S_z$ &\ \ Eigenvalue \\
		\hline \hline	
		1 &\ \ $|0,0\rangle$ &\ \ 0  &\ \ 0 &\ \  0 &\ \  0  \\
		\hline
		2 &\ \ $|0,\uparrow\rangle$&\ \ 1 &\ \ 1/2 & \ \ 1/2 &\ \  $-\Delta/2-\mu$ \\
		3 &\ \ $|\uparrow,0\rangle$&\ \ 1 &\ \ 1/2 & \ \ 1/2 &\ \  $\Delta/2-\mu$ \\
		4 &\ \ $|0,\downarrow\rangle$&\ \ 1 &\ \ 1/2 & \ \ -1/2 &\ \  $-\Delta/2-\mu$ \\
		5 &\ \ $|\downarrow,0\rangle$&\ \ 1 &\ \ 1/2 & \ \ -1/2 &\ \  $\Delta/2-\mu$ \\		
		\hline
		6 &\ \ $|\uparrow,\uparrow\rangle$ &\ \ 2 &\ \ 1 &\ \ 1 &\ \  $U-3J-2\mu$ \\ 
		7 &\ \ $\big(|\uparrow,\downarrow\rangle + |\downarrow,\uparrow\rangle\big)/\sqrt{2}$ &\ \ 2 &\ \ 1 &\ \ 0 &\ \  $U-3J-2\mu$ \\ 
		8 &\ \ $|\downarrow,\downarrow\rangle$ &\ \ 2 &\ \ 1 &\ \ -1 &\ \  $U-3J-2\mu$ \\
		9 &\ \ $\frac{a+b}{\sqrt{(a+b)^2+J^2}}|\uparrow\downarrow,0\rangle - \frac{J}{\sqrt{(a+b)^2+J^2}} |0,\downarrow\uparrow\rangle$ &\ \ 2 &\ \ 0 &\ \ 0 &\ \  $U-\sqrt{J^2+\Delta^2}-2\mu$ \\ 
		10 &\ \ $\big(|\uparrow,\downarrow\rangle - |\downarrow,\uparrow\rangle\big)/\sqrt{2}$ &\ \ 2 &\ \ 0 &\ \ 0 &\ \  $U-J-2\mu$ \\ 
		11 &\ \ $\frac{J}{\sqrt{(a+b)^2+J^2}}|\uparrow\downarrow,0\rangle + \frac{a+b}{\sqrt{(a+b)^2+J^2}} |0,\downarrow\uparrow\rangle$ &\ \ 2 &\ \ 0 &\ \ 0 &\ \  $U+\sqrt{J^2+\Delta^2}-2\mu$ \\ 
		\hline
		12 &\ \ $|\uparrow\downarrow,\uparrow\rangle$&\ \ 3 &\ \ 1/2 & \ \ 1/2 &\ \  $3U-5J+\Delta/2-3\mu$ \\
		
		13 &\ \ $|\uparrow,\uparrow\downarrow\rangle$&\ \ 3 &\ \ 1/2 & \ \ 1/2 &\ \  $3U-5J-\Delta/2-3\mu$ \\
		14 &\ \ $|\uparrow\downarrow,\downarrow\rangle$&\ \ 3 &\ \ 1/2 & \ \ -1/2 &\ \  $3U-5J+\Delta/2-3\mu$ \\
		15 &\ \ $|\downarrow,\uparrow\downarrow\rangle$&\ \ 3 &\ \ 1/2 & \ \ -1/2 &\ \  $3U-5J-\Delta/2-3\mu$ \\
		\hline	   
		16 &\ \ $|\uparrow \downarrow, \uparrow \downarrow \rangle$ &\ \ 4 &\ \ 0 &\ \ 0 &\ \ $6U-10J-4\mu$ \\
		\hline \hline
		
	\end{tabular}
	\caption{Eigenstates and eigenvalues of Eq.~(1) with $U'=U-2J$, $\epsilon_1=\Delta/2$, and $\epsilon_2=-\Delta/2$. $a\equiv-\Delta$ and $b\equiv\sqrt{J^2+\Delta^2}$. The first entry in a ket of an eigenstate is the state of orbital-1 and the second is of orbital-2. }
	\label{table_s3}
\end{table*}

Having evidenced that the sign of $-(\partial \mathcal{J}_\mathrm{s} / \partial \mu) d\mu$ is influenced by $J$, we now consider Eq.~(\ref{eq1}) with $U'=U-2J$ for $H_\mathrm{loc}$. In this case, only $\mathrm{SU(2)}$ symmetry of spin is retained. Using eigenstates and eigenvalues listed in Table~\ref{table_s2} and Table~\ref{table_s3}, we get the following relations for $\mathcal{J}_\mathrm{s}$:
\begin{align}
	\mathcal{J}_\mathrm{s} &= 
	\frac{V^2}{2} \Bigg( \frac{2}{ E_0^{|4,0\rangle}} -\frac{1}{ E_0^{|2,1\rangle}} + \frac{2}{ E_0^{|2,0\rangle}} + \frac{1}{ E_1^{|2,0\rangle}} \Bigg) \;\; \mathrm{for}~\Delta = 0, \label{eqS22}\\
	\mathcal{J}_\mathrm{s} &= 
	\frac{V^2}{2} \Bigg( \frac{2}{ E_0^{|4,0\rangle}} -\frac{1}{ E_0^{|2,1\rangle}} + \frac{2J^2}{ (a+b)^2+J^2} \frac{1}{ E_0^{|2,0\rangle}} \nonumber \\ &+ \frac{1}{ E_1^{|2,0\rangle} } + \frac{2(a+b)^2}{ (a+b)^2+J^2} \frac{1}{ E_2^{|2,0\rangle}}  \Bigg) \;\; \mathrm{for}~\Delta > 0, \label{eqS23}
\end{align}
where $a\equiv-\Delta$ and $b\equiv \sqrt{J^2+\Delta^2}$.
Applying the same procedure used for getting Eq.~(\ref{eqS19})  for $J>0$ results in 
\begin{align}
	-&\Big( \frac{\partial \mathcal{J}_\mathrm{s}}{\partial \mu} \Big) d\mu  \approx -\frac{1}{2}\frac{V^2}{\big(E^{|2,1\rangle}_0\big)^2 } d\mu \;\; \mathrm{for}~\Delta = 0, \label{eqS24} \\
	-&\Big( \frac{\partial \mathcal{J}_\mathrm{s}}{\partial \mu} \Big) d\mu \approx -\frac{1}{2} \frac{V^2}{ \big(E_0^{|2,1\rangle} \big)^2} d\mu  \nonumber \\ & + \frac{1}{\big\{ \sqrt{1+(\Delta/J)^2}-\Delta/J \big\}^2+1} \frac{V^2}{ \big(E_0^{|2,0\rangle} \big)^2} d\mu \;\; \mathrm{for}~\Delta > 0. \label{eqS25}
\end{align}
The Eq.~(\ref{eqS25}) above is the same as Eq.~(\ref{eq3}).

Finally, we briefly discuss the case of $M=2$ and $n_0=2$ with Eq.~(\ref{eq1}) being $H_\mathrm{loc}$. We set $|\phi_0\rangle = |\uparrow,\uparrow \rangle$ and $\alpha=3$.
Using eigenstates and eigenvalues listed in Table~\ref{table_s2} and Table~\ref{table_s3}, we arrive at
\begin{align}
	\mathcal{J}_\mathrm{s} & = V^2 \Bigg( \frac{1}{ E_0^{|3,1/2\rangle}} + \frac{1}{ E_0^{|1,1/2\rangle}}  \Bigg) \;\; \mathrm{for}~\Delta = 0, \label{eqS26}\\
	\mathcal{J}_\mathrm{s} &= \frac{V^2}{2} \Bigg( \frac{1}{ E_0^{|3,1/2\rangle}} + \frac{1}{ E_1^{|3,1/2\rangle}} \nonumber \\ & + \frac{1}{ E_0^{|1,1/2\rangle}} + \frac{1}{ E_1^{|1,1/2\rangle}} \Bigg) \;\; \mathrm{for}~\Delta > 0, \label{eqS27}
\end{align}
where $E_k^{|N,S\rangle}$ ($k \in \{0,1,...\}$) in this case denotes the excitation energy from $|2,1\rangle$ to the excited atomic multiplet $|N,S \rangle$. The change of $\mathcal{J}_\mathrm{s}$ under $\mu \rightarrow \mu- d\mu$ is given by
\begin{align}
	-\Big( \frac{\partial \mathcal{J}_\mathrm{s}}{\partial \mu} \Big) d\mu &=  V^2 \Bigg\{ \frac{1}{\big(E^{|1,1/2\rangle}_0\big)^2} - \frac{1}{\big(E^{|3,1/2\rangle}_0\big)^2} \Bigg\} d\mu \;\; \mathrm{for}~\Delta = 0, \label{eqS28} \\
	-\Big( \frac{\partial \mathcal{J}_\mathrm{s}}{\partial \mu} \Big) d\mu &= V^2 \Bigg\{ \frac{1}{\big(E^{|1,1/2\rangle}_0\big)^2} + \frac{1}{\big(E^{|1,1/2\rangle}_1\big)^2} - \frac{1}{\big(E^{|3,1/2\rangle}_0\big)^2}  \nonumber \\ &- \frac{1}{\big(E^{|3,1/2\rangle}_1\big)^2} \Bigg\} d\mu \;\; \mathrm{for}~\Delta > 0. \label{eqS29}
\end{align}
As $E^{|1,1/2\rangle}_k < E^{|3,1/2\rangle}_k$ for $p > 0$, Eqs.~(\ref{eqS28}--\ref{eqS29}) are always positive. Note that $-(\partial \mathcal{J}_\mathrm{s} / \partial \mu) d\mu \geq 0$ irrespective of $J$ and $\Delta$ for $p \geq 0$, which is in sharp contrast to the case of $n_0=3$.

\section{$\mathcal{J}_\mathrm{s}$ and $R_\mathrm{s}^{-1}$ as a function of $p$ for two-orbital models on a Bethe lattice} \label{appendix_E}

\begin{figure*} [!htbp]
	\includegraphics[width=0.83\textwidth, angle=0]{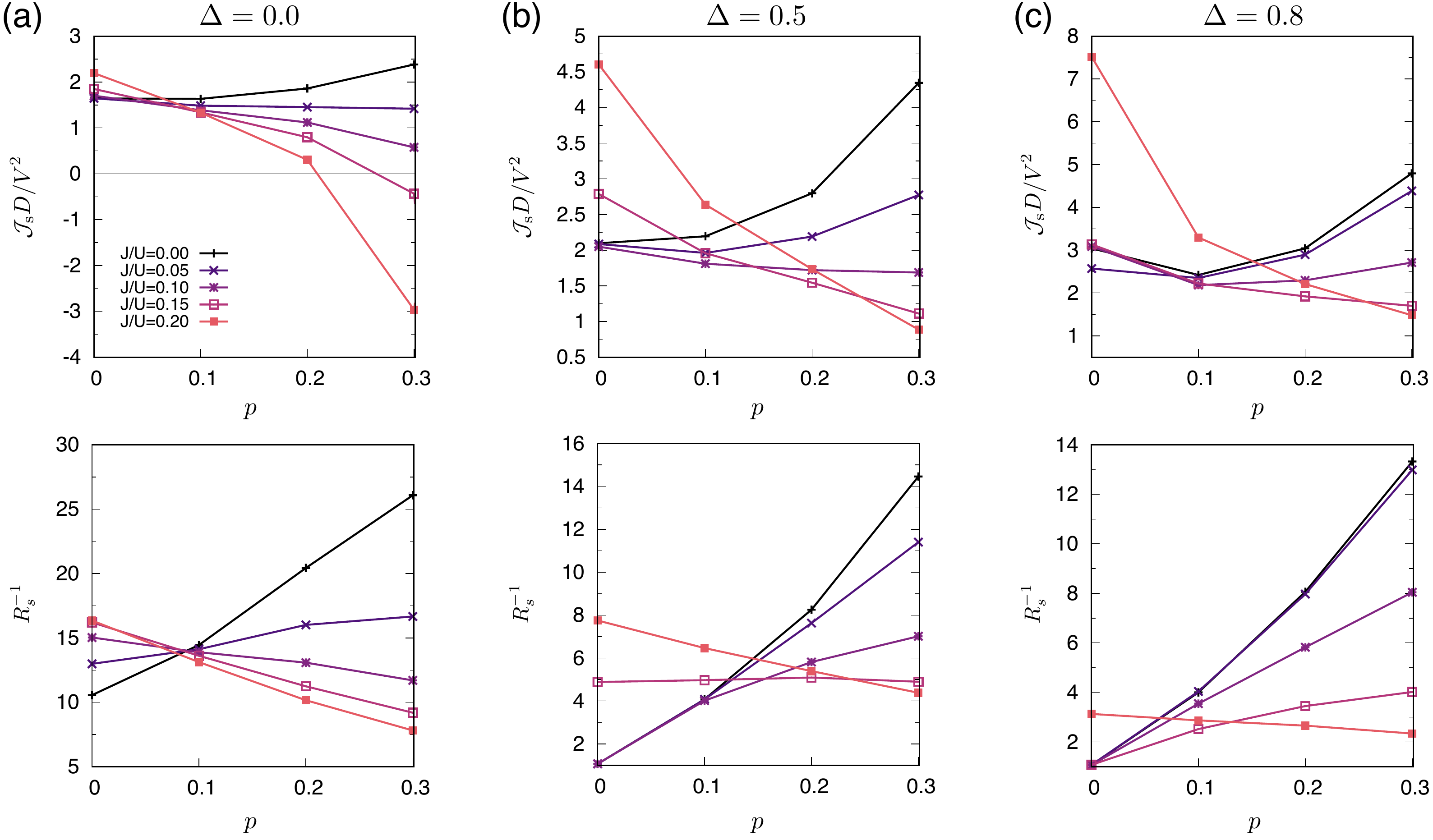}
	\caption{$\mathcal{J}_\mathrm{s}$ (multiplied by $D/V^2$) and $R_\mathrm{s}^{-1}$ as a function of $p$ for generic two-orbital models on a Bethe lattice at three different values of $\Delta$: $\Delta=0$ (leftmost), 0.5 (middle), and 0.8 (rightmost). $U=2.5$ and $n=3-p$ for all the cases. We used $E_k^{|N,S\rangle}$s obtained from the DMFT solutions for the evaluation of $\mathcal{J}_\mathrm{s}$.}
	\label{fig9}
\end{figure*}

Figure~\ref{fig9} displays $\mathcal{J}_\mathrm{s}$ and $R_\mathrm{s}^{-1}$ as a function of $p$ for two-orbital models on a Bethe lattice with $\Delta \geq 0$. Although $\mathcal{J}_\mathrm{s}$ is a bare Kondo coupling which will be scaled via renormalization group flow, the behavior of $\mathcal{J}_\mathrm{s}$ as a function of $p$ is qualitatively consistent with that of $R_\mathrm{s}^{-1}$.

\section{$\partial R_\mathrm{s}^{-1}/ \partial p$ and its relation to $\mathcal{J}_\mathrm{s}$ for the cases of $p<0$ and $p = 1 - \epsilon$} \label{appendix_F}

Our criterion based on the sign of $\partial R_s^{-1} / \partial p$ cannot distinguish Mott and Hund correlations for the negative $p$ ($p<0$) and very large $p$ of $p = 1-\epsilon$ ($\epsilon$ is an arbitrarily small positive number) by which the system is close to half-filling. To understand these cases, let us focus on a two-degenerate-orbital model with $U'=U-2J$ in Eq.~(\ref{eqS2}).

\subsection{The case of $p<0$.} 

This case is exactly the same as the ``electron doping" $x$ ($x>0$) to a system of $n=M+1=3$ electron filling. For this filling, Kondo coupling for the spin degree of freedom, $\mathcal{J}_\mathrm{s}$, is given by Eq.~(\ref{eqS13}). We now consider response of $\mathcal{J}_\mathrm{s}$ due to electron doping $x$. To mimic the effect of a small increase in $x$, let us consider a situation where $\mu$ is slightly increased by $d\mu$ ($d\mu >0$), i.e., $\mu \rightarrow \mu + d\mu$. As a consequence, $\mathcal{J}_\mathrm{s} \rightarrow \mathcal{J}_\mathrm{s} + \big( \partial \mathcal{J}_\mathrm{s} / \partial \mu \big) d\mu $.
Thus, the concomitant change in $\mathcal{J}_\mathrm{s}$ is given by:
\begin{align}
	\Big( \frac{\partial \mathcal{J}_\mathrm{s}}{\partial \mu} \Big) d\mu &= \frac{V^2}{2} \Bigg\{ \frac{2}{ \big(E_0^{|4,0\rangle}\big)^2 } +\frac{1}{ \big(E_0^{|2,1\rangle}\big)^2 } - \frac{3}{ \big(E_0^{|2,0\rangle}\big)^2 } \Bigg\} d\mu. \label{eqS30}
\end{align}

When $J=0$, $E_0^{|2,1\rangle}$ is equal to $E_0^{|2,0\rangle}$. Furthermore, since $E_0^{|4,0\rangle}$ is smaller than $E_0^{|2,1\rangle}$ for the electron doped side, Eq.~(\ref{eqS30}) is positive. This means that Kondo screening for spin becomes more effective by electron doping, and thereby correlation strength is reduced. Thus, $\partial R_\mathrm{s}^{-1} / \partial x < 0$. This conclusion is actually consistent with our physical intuition that Mott physics is weakened by doping an integer-filled system. 

When $J$ is large, $E_0^{|2,1\rangle}$ is much smaller than $E_0^{|2,0\rangle}$. This observation leads us to the following expression.  
\begin{align}
	\Big( \frac{\partial \mathcal{J}_\mathrm{s}}{\partial \mu} \Big) d\mu & \simeq \frac{V^2}{2} \Bigg\{ \frac{2}{ \big(E_0^{|4,0\rangle}\big)^2 } +\frac{1}{ \big(E_0^{|2,1\rangle}\big)^2 } \Bigg\} d\mu. \label{eq4}
\end{align}
As in the case of $J=0$, $\mathcal{J}_\mathrm{s}$ also increases by electron doping. Thus, we expect $\partial R_\mathrm{s}^{-1} / \partial x < 0$ even for the case of large $J$. 

To conclude our discussion on the case of $p<0$, we can confirm from the analysis of $\mathcal{J}_\mathrm{s}$ that spin screening becomes more effective upon doping no matter how large $J$ is. Physically, this is because the case of $p<0$ (or, equivalently electron filling of $n=M+1=3+p$) is the doping by which the ``Hund fluctuations" (ferromagnetic charge fluctuations between the dominant atomic multiplets and higher-spin ones in a neighboring charge subspace, i.e., the term containing $E_0^{|2,1\rangle}$) are weakened. Thus, our criterion based on the sign of $\partial R_s^{-1} / \partial p$ is not applicable to this case.

\subsection{The case of $p = 1-\varepsilon$ ($\varepsilon$ is an arbitrarily small positive number).}

This case is exactly the same as the ``electron doping" $x$ ($x>0$) to a system of $n=M=2$ electron filling (half filling). Kondo coupling for spin is given by Eq.~(\ref{eqS26}). To mimic the effect of a small increase in $x$, let us consider a situation where $\mu$ is slightly increased by $d\mu$ ($d\mu >0$), i.e., $\mu \rightarrow \mu + d\mu$. As a consequence, $\mathcal{J}_\mathrm{s} \rightarrow \mathcal{J}_s + \big( \partial \mathcal{J}_\mathrm{s} / \partial \mu \big) d\mu $.
Thus, the concomitant change in $\mathcal{J}_s$ is given by:
\begin{align}
	\Big( \frac{\partial \mathcal{J}_\mathrm{s}}{\partial \mu} \Big) d\mu &=  V^2 \Bigg\{ \frac{1}{\big(E_0^{|3,1/2\rangle}\big)^2} - \frac{1}{\big(E_0^{|1,1/2\rangle}\big)^2} \Bigg\}   d\mu. \label{eqS32}
\end{align}
Since $E_0^{|3,1/2\rangle} < E_0^{|1,1/2\rangle}$ on the electron doped side, Eq.~(\ref{eqS32}) is always positive irrespective of $J$.  This is because the role of doping in this case is to reduce the weight of the high-spin multiplet, namely $|2,1\rangle$ which is responsible for realizing Hund metal physics. Our criterion measures how large the effect of the high-spin ``half-filled" multiplet (i.e., $|2,1\rangle$ for $M=2$) is on the low-energy physics while the system is far away from half fillng. Since $|2,1\rangle$ already dominates the atomic states, it is not surprising at all that our proposal is inapplicable to this case.


%

\end{document}